\documentclass{ar-1col}

  \usepackage{color,colortbl}
  \definecolor{colurl}{rgb}{0.2,0.2,1}
  \definecolor{colcite}{rgb}{0.2,0.2,1}
  \definecolor{collink}{rgb}{0.5,0.,0.5}
  \definecolor{colsec}{rgb}{0.5,0.,0.5}
  \definecolor{colhead}{rgb}{0.6,0.6,0.6}
  \definecolor{coltab1}{cmyk}{0.03,0.03,0.12,0}
  \definecolor{coltab2}{cmyk}{0.06,0.02,0.04,0}
  \definecolor{colunc}{rgb}{0.5,0.5,0.5}
  \usepackage[colorlinks=true,
              breaklinks=true,
              pdfstartview=FitV, 
              linkcolor=collink, 
              filecolor=collink,
              citecolor=colcite, 
              urlcolor=colurl, 
              backref=true,
              setpagesize=false]{hyperref}
  \usepackage{natbib}
  \usepackage{amssymb}
  \usepackage{tabularx}

  \usepackage{savesym}
  \savesymbol{marginnote}
  \usepackage{pdfcomment}
  \restoresymbol{origmarginnote}{marginnote}

  \jname{Xxxx. Xxx. Xxx. Xxx.}
  \jvol{AA}
  \jyear{YYYY}
  \doi{10.1146/((please add article doi))}

  \graphicspath{{Figures/}{.}}
  \newcommand{\sectbox}[1]{\section{#1}\vspace{-18pt}%
                           \addcontentsline{toc}{subsection}{\fbox{BOX: #1}}}
  \newcommand{\sectboxns}[1]{\section{#1}%
                           \addcontentsline{toc}{subsection}{\fbox{BOX: #1}}}
  
  \newcommand{\subsubsectbox}[1]{\subsubsection{#1}}
  
  \newcommand{\mmic}{\;\mu\rm m}
  \newcommand{\mic}{$\mu\rm m$}
  \newcommand{\etc}{{\it etc.}}
  \newcommand{\eg}{{\it e.g.}}
  
  \newcommand{\cf}{{cf.}}
  \newcommand{\insitu}{{\it in situ}}
  \newcommand{\apriori}{{\it a priori}}
  \newcommand{\sms}[1]{{\mbox{{\scriptsize #1}}}}
  \newcommand{\textitem}[1]{\textbf{(#1)}}
  \newcommand{\citext}[1]{{\it#1}}

  \newcommand{\unc}[1]{\textcolor{colunc}{#1}}
  
  \newcommand{\Rv}{$R_{V}$}
  \newcommand{\mRv}{R_{V}}
  \newcommand{\Av}{$A_{V}$}
  \newcommand{\mAv}{A_{V}}
  
  \newcommand{\mAb}{A_{B}}
  \newcommand{\DG}{Z_\sms{dust}}
  \newcommand{\mN}[1]{N(\mbox{#1})}
  \newcommand{\N}[1]{$N(\mbox{#1})$}

  \newcommand{\mSurf}[1]{\Sigma_\sms{#1}}
  \newcommand{\Surf}[1]{$\Sigma_\sms{#1}$}
  \newcommand{\Lnu}[1]{L_\nu({#1})}
  \newcommand{\Lint}[1]{L_{#1}}

  \newcommand{\hi}{H$\,${\sc i}}
  \newcommand{\hii}{H$\,${\sc ii}}
  \newcommand{\hmol}{H$_2$}
  \newcommand{\HAC}{a-C(:H)}

  \newcommand{\cii}{C$\,${\sc ii}}
  
  \newcommand{\ha}{H$\alpha$}
  \newcommand{\bra}{Br$\alpha$}

  \newcommand{\neiii}{Ne$\,${\sc iii}}
  \newcommand{\neii}{Ne$\,${\sc ii}}

  \newcommand{\oi}{O$\,${\sc i}}

  \newcommand{\hiline}{[\hi]$_{21\,\rm cm}$}

  \newcommand{\ciiline}{[\cii]$_{158\mu\rm m}$}

  \newcommand{\neiiiline}{[\neiii]$_{15.56\mu\rm m}$}
  \newcommand{\neiiline}{[\neii]$_{12.81\mu\rm m}$}

  \newcommand{\oiline}{[\oi]$_{63\mu\rm m}$}

  \newcommand{\COio}{$^{12}$CO(J$=$1$\rightarrow$0)$_{2.6\rm mm}$}

  \newcommand{\ngc}[1]{NGC$\,$#1}
  \newcommand{\M}[1]{M$\,$#1}

  \newcommand{\cena}{Cen~A}
  \newcommand{\xxxdor}{30$\,$Dor}
  \newcommand{\sbs}{SBS$\,$0335-052}
  \newcommand{\izw}{I$\,$Zw$\,$18}  

  \newcommand{\reffig}[1]{\hyperref[#1]{Figure~\ref{#1}}}
  \newcommand{\refpanel}[2]{\hyperref[#1]{Figure~\ref{#1}-#2}}
  \newcommand{\reftab}[1]{\hyperref[#1]{Table~\ref{#1}}}
  \newcommand{\refsec}[1]{\hyperref[#1]{Section~\ref{#1}}}
  \newcommand{\refsecs}[2]{\hyperref[#1]{Sections~\ref{#1}} and
                           \ref{#2}}
  \newcommand{\refeq}[1]{\hyperref[#1]{Equation~(\ref{#1})}}
  \newcommand{\refeqp}[1]{(\hyperref[#1]{Equation~\ref{#1}})}

  \newcommand{\iras}{{\it IRAS}}
  \newcommand{\iso}{{\it ISO}}

  \newcommand{\spitz}{{\it Spitzer}}
  \newcommand{\akari}{{\it AKARI}}
  \newcommand{\hersc}{{\it Herschel}}
  \newcommand{\planck}{{\it Planck}}
  \newcommand{\alma}{{\it ALMA}}
  
  \newcommand{\jwst}{{\it JWST}}
  \newcommand{\spica}{{\it SPICA}}
  
  \newcommand{\cobe}{{\it COBE}}
  \newcommand{\wmap}{{\it WMAP}}
  \newcommand{\luvoir}{{\it LUVOIR}}

  \defcitealias{calzetti94}{C94}
  \defcitealias{cardelli89}{C89}
  \defcitealias{conroy10}{C10}
  \defcitealias{draine07}{DL07}
  \defcitealias{gordon03}{G03}
  \defcitealias{galliano08a}{G08a}
  \defcitealias{galliano08b}{G08b}
  \defcitealias{galliano18}{G18}
  \defcitealias{kohler15}{K15}
  \defcitealias{jones17}{J17}
  \defcitealias{smith07}{S07}
  \defcitealias{planck-collaboration14c}
               {\planck~\citeyearpar{planck-collaboration14c}}
  \defcitealias{planck-collaboration15c}
               {\planck~\citeyearpar{planck-collaboration15c}}
  \defcitealias{planck-collaboration11}
               {\planck~\citeyearpar{planck-collaboration11}}

  \newcommand{\mytip}[2]{\pdftooltip{\textcolor{collink}{#1}}{#2}}
  \newcommand{\hRT}{\mytip{RT}{Radiative transfer}}
  \newcommand{\hAGB}{\mytip{AGB}{Asymptotic giant branch}}
  \newcommand{\hAGN}{\mytip{AGN}{Active galactic nuclues}}
  \newcommand{\hakari}{\mytip{\akari}{Akari space telescope 
                                      (1.7-180 microns; 2006-2011)}}
  \newcommand{\hALMA}{\mytip{\alma}{Atacama large millimeter array 
                                    (300 microns-1 cm; 2011)}}
  \newcommand{\hAME}{\mytip{AME}{Anomalous microwave emission}}
  \newcommand{\hBCD}{\mytip{BCD}{Blue compact dwarf}}
  \newcommand{\hbeta}{\mytip{$\beta$}{Emissivity index (power-law index of 
                                      a modified black body)}}
  \newcommand{\hbeff}{\mytip{$\beta_\sms{eff}$}{Effective emissivity index 
                                                (fitted value of beta)}}
  \newcommand{\hcobe}{\mytip{\cobe}{Cosmic Background Explorer 
                                    (12-5000 microns; 1989-1993)}}
  \newcommand{\hDG}{\mytip{$\DG$}{Dust-to-gas mass ratio}}
  \newcommand{\hDIB}{\mytip{DIB}{Diffuse interstellar band}}
  \newcommand{\hDLA}{\mytip{DLA}{Damped Lyman-alpha systems}}
  \newcommand{\hDNM}{\mytip{DNM}{Diffuse neutral medium}}
  \newcommand{\hERE}{\mytip{ERE}{Extended red emission}}
  \newcommand{\hETG}{\mytip{ETG}{Early-type (evolved) galaxy}}
  
  \newcommand{\hFIR}{\mytip{FIR}{Far-infrared}}
  \newcommand{\hFUV}{\mytip{FUV}{Far-ultraviolet}}
  
  \newcommand{\hHAC}{\mytip{a-C(:H)}{(hydrogenated) amorphous carbon}}
  \newcommand{\hhersc}{\mytip{\hersc}{Herschel space observatory 
                                      (55-672 microns; 2009-2013)}}
  \newcommand{\hI}{\mytip{$I$-band}{880 nm}}
  \newcommand{\hIMF}{\mytip{IMF}{Initial mass function}}
  \newcommand{\hIR}{\mytip{IR}{Infrared}}
  \newcommand{\hiras}{\mytip{\iras}{Infrared astronomical satellite 
                                   (12-100 microns; 1983)}}
  \newcommand{\hISM}{\mytip{ISM}{Interstellar medium}}
  \newcommand{\hiso}{\mytip{\iso}{Infrared Space Telescope 
                                  (5-210 microns; 1995-1998)}}

  \newcommand{\hISRF}{\mytip{ISRF}{Interstellar radiation field}}
  \newcommand{\hjwst}{\mytip{\jwst}{James Webb Space Telescope 
                                   (0.6-27 microns; 2018-)}}

  \newcommand{\hLMC}{\mytip{LMC}{Large Magellanic cloud}}
  \newcommand{\hLint}[1]{\mytip{$\Lint{#1}$}{Integrated luminosity of the #1 
                                             micron feature (Lsun)}}
  \newcommand{\hLIRG}{\mytip{LIRG}{Luminous infrared galaxy 
                                   (1.E11 Lsun < L(IR) < 1.E12 Lsun)}}
  \newcommand{\hLnu}[1]{\mytip{$\Lnu{#1}$}{Monochromatic luminosity at #1 
                                         microns (Lsun/Hz)}}
  \newcommand{\hluvoir}{\mytip{\luvoir}{UV-NIR space telescope 
                                        (in preparation)}}
  \newcommand{\hMBB}{\mytip{MBB}{Modified black body}}
  \newcommand{\hmbeta}{\mytip{\beta}{Emissivity index (power-law index of 
                                     a modified black body)}}
  \newcommand{\hmDG}{\mytip{\DG}{Dust-to-gas mass ratio}}
  \newcommand{\hmet}{\mytip{$Z$}{Metallicity}}
  \newcommand{\hMIR}{\mytip{MIR}{Mid-infrared}}
  \newcommand{\hmLnu}[1]{\mytip{\Lnu{#1}}{Monochromatic luminosity at #1 
                                          microns (Lsun/Hz)}}
  \newcommand{\hMW}{\mytip{MW}{Milky Way}}
  \newcommand{\hNIR}{\mytip{NIR}{Near-infrared}}
  \newcommand{\hNUV}{\mytip{NUV}{Near-ultraviolet}}
  \newcommand{\hPAH}{\mytip{PAH}{Polycyclic aromatic hydrocarbon}}
  \newcommand{\hPE}{\mytip{PE}{Photoelectric}}
  
  \newcommand{\hPDR}{\mytip{PDR}{Photodissociation region}}
  \newcommand{\hplanck}{\mytip{\planck}{Planck space observatory 
                                        (300 microns-1 cm; 2009-2013)}}
  \newcommand{\hSED}{\mytip{SED}{Spectral energy distribution}}
  \newcommand{\hSFR}{\mytip{SFR}{Star formation rate}}
  
  \newcommand{\hSMC}{\mytip{SMC}{Small Magellanic cloud}}
  \newcommand{\hSN}{\mytip{SN}{Supernova}}
  \newcommand{\hSNIa}{\mytip{SN$\,$Ia}{Type Ia supernova}}
  \newcommand{\hSNII}{\mytip{SN$\,$II}{Type II supernova}}
  \newcommand{\hspica}{\mytip{\spica}{MIR–FIR space telescope (12−210 microns; 
                                      launch in 2025)}}
  \newcommand{\hspitz}{\mytip{\spitz}{Spitzer space telescope 
                                      (3-160 microns; 2003-2009)}}
  \newcommand{\hsSFR}{\mytip{sSFR}{Specific star formation rate (SFR/M*)}}
  \newcommand{\hSSC}{\mytip{SSC}{Super Star Cluster}}
  \newcommand{\hsubmm}{\mytip{submm}{Submillimeter}}
  \newcommand{\hTIR}{\mytip{TIR}{Total infrared}}
  \newcommand{\hU}{\mytip{$U$}{Interstellar radiation field intensity}}
  \newcommand{\hmU}{\mytip{U}{Interstellar radiation field intensity}}
  \newcommand{\hUIB}{\mytip{UIB}{Unidentified infrared band}}
  \newcommand{\hULIRG}{\mytip{ULIRG}{Ultraluminous infrared galaxy 
                                     (L(IR) > 1.E12 Lsun)}}
  \newcommand{\hUV}{\mytip{UV}{Ultraviolet}}
  \newcommand{\hV}{\mytip{$V$-band}{560 nm}}
  \newcommand{\hwmap}{\mytip{\wmap}{Wilkinson Microwave Anisotropy Probe
                                    (3.2-13 mm; 2001-2010)}}

\begin{document}

  \markboth{Galliano, Galametz \&\ Jones}
           {The Interstellar Dust Properties of Nearby Galaxies}

  \title{The Interstellar Dust Properties of Nearby Galaxies}

  \author{Fr\'ed\'eric Galliano,$^{1,2}$ Maud Galametz,$^{1,2}$ 
          and Anthony P.\ Jones$^3$
    \affil{$^1$IRFU, CEA, Universit\'e Paris-Saclay, F-91191 Gif-sur-Yvette, 
           France \\
           email: frederic.galliano@cea.fr, maud.galametz@cea.fr}
    \affil{$^2$Universit\'e Paris-Diderot, AIM, Sorbonne Paris Cit\'e, CEA, 
             CNRS, F-91191 Gif-sur-Yvette, France}
    \affil{$^3$Institut d'Astrophysique Spatiale, CNRS, Univ. Paris-Sud, 
           Universit\'e Paris-Saclay, B\^at.\ 121, 91405 Orsay, France;
           email: anthony.jones@ias-u.psud.fr}}

  \begin{abstract}
    This article gives an overview of the constitution, physical 
    conditions and observables of dust in the interstellar medium of nearby 
    galaxies.
	We first review the macroscopic, spatial distribution of dust in these 
	objects, and its consequence on our ability to study grain physics.
	We also discuss the possibility to use dust tracers as diagnostic tools.
	We then survey our current understanding of the microscopic, intrinsic
	properties of dust in different environments, derived from different 
	observables: emission, extinction, polarization, depletions, over the whole
	electromagnetic spectrum.
    Finally, we summarize the clues of grain evolution, evidenced either on 
    local scales or over cosmic time.
    We put in perspective the different evolution scenarios.
    We attempt a comprehensive presentation of the main observational
    constraints, analysis methods and modelling frameworks of the distinct 
    processes. 
	We do not cover the dust properties of the Milky Way and distant 
	galaxies, nor circumstellar or active galactic nucleus torus dust. 
  \end{abstract}

  \maketitle


  \section{INTRODUCTION}
  \label{sec:intro}

  \subsection{The Interstellar Dust: a Key Galaxy Component}
  \label{sec:introdust}

Interstellar grains are solid particles of sizes 
$0.3\;{\rm nm}\lesssim r\lesssim0.3\mmic$, made of heavy elements 
(mainly O, C, Si, Mg, Fe) available in the InterStellar Medium (\hISM).
They appear to be rather uniformly mixed with the gas.
Although accounting for $\lesssim1\,\%$ of the \hISM\ mass, they have a 
radical impact on galaxies, as they scatter and absorb starlight.
In normal disk galaxies, they re-radiate in the InfraRed (\hIR) about $30\,\%$ 
of the stellar power, and up to $99\,\%$ in ultraluminous \hIR\ galaxies 
\citep[\eg][]{clements96}.
In addition, they are responsible for the heating of the gas in PhotoDissociation Regions (\hPDR),
by photoelectric effect \citep[{\hPE};][]{draine78}.
They are also catalysts of numerous chemical reactions, including the
formation of the most abundant molecule in the Universe, \hmol\ 
\citep{gould63}.

\begin{figure}[h]
  \includegraphics[width=1.23\textwidth]{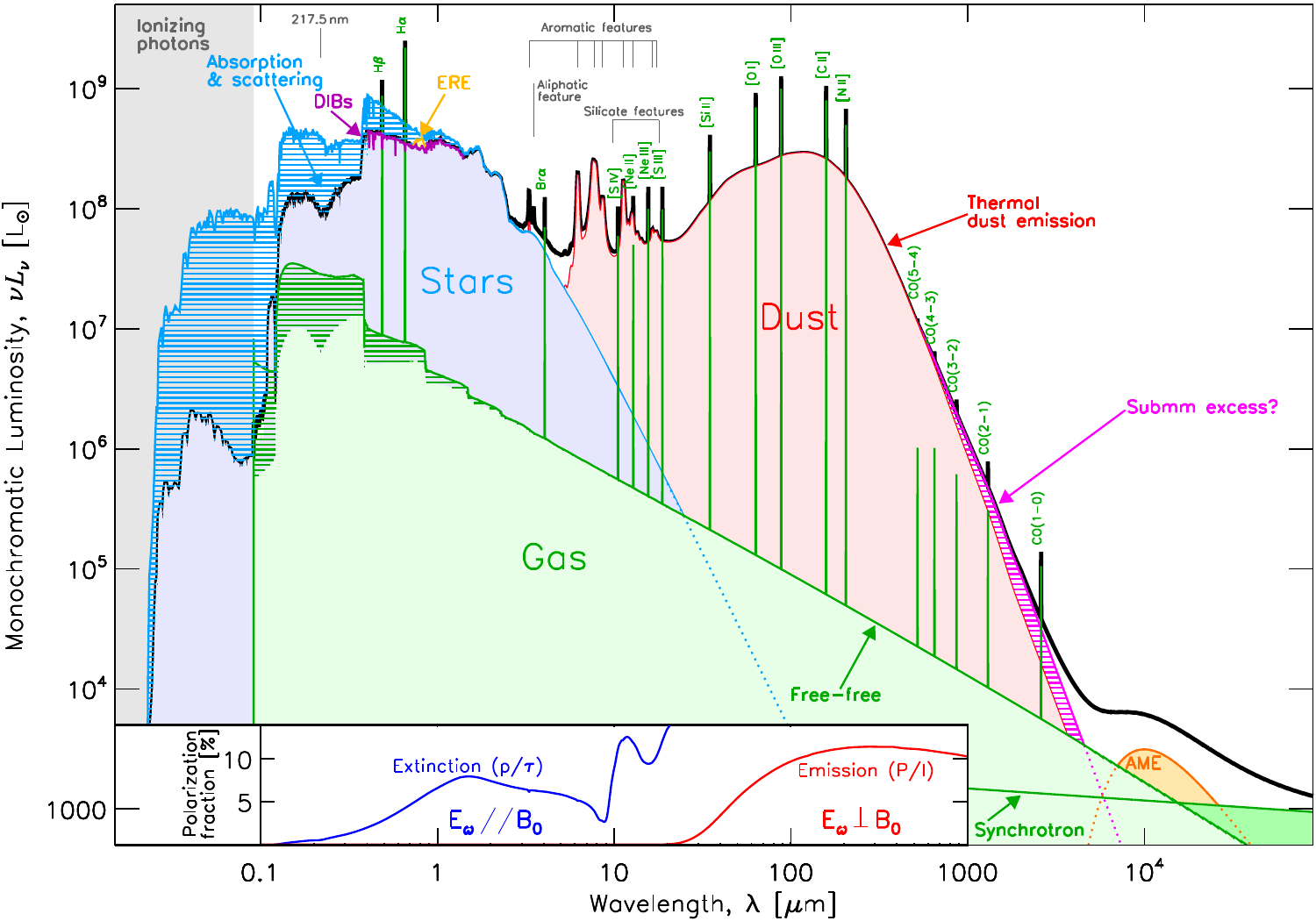}
  \caption{{\sl Spectral energy distribution (\hSED) of a typical late-type 
           galaxy.}
           The blue hatched area shows the power absorbed by dust.
           We show typical Diffuse Interstellar Band (\hDIB), Extended Red
           Emission (\hERE) and Anomalous Microwave Emission (\hAME) spectra,
           with the most relevant gas lines.
           The inset shows the model D of \citet[][$G_0=100$]{guillet17}.}
  \label{fig:dustobs}
\end{figure}

A detailed knowledge of the grain properties is crucial to study the lifecycle 
of the \hISM\ and galaxy evolution, as it is needed to:
\textitem{1}~unredden \hUV-visible observations;
\textitem{2}~study deeply embedded regions;
\textitem{3}~build reliable diagnostics of the physical conditions and of
  the evolutionary stage of a galaxy or a star forming region;
\textitem{4}~provide accurate prescriptions in photoionization and
  photodissociation models, and simulations of the star formation process.
However, as we will show in this review, there remains several uncertainties 
about the grain properties and their evolution.
Dust physics is characterized by the great complexity of its object.
The number of ways to combine elements to build interstellar solids
is virtually limitless and has consequences on the longevity of the particle
and its observables.
The progress in this field is thus mainly driven by empirical 
constraints: 
observations over the whole electromagnetic spectrum (\reffig{fig:dustobs});
and laboratory experiments on dust analogs.

\begin{textbox}[h]
  \sectbox{WHAT ARE THE ``DUST PROPERTIES''?}%
  \subsubsectbox{Dust mixture constitution}%
    The constitution of a grain mixture is defined by: 
    \textitem{1}~the chemical composition of the bulk material and its 
    stoichiometry;
    \textitem{2}~the structure of the grains (crystalline, amorphous, 
      porous, aggregated, \etc);
    \textitem{3}~the presence of heterogeneous inclusions;
    \textitem{4}~the presence of organic and/or icy mantles;
    \textitem{5}~the shape of the grains;
    \textitem{6}~their size distribution;
    \textitem{7}~their abundance relative to the gas.
  \subsubsectbox{Dust physical conditions}%
    A dust mixture, with a given constitution, can experience
    different physical conditions:
    \textitem{1}~thermal excitation of the grains, due to radiative heating 
    (equilibrium or stochastic), or to collisional heating in a hot plasma;
    \textitem{2}~grain charging by exchange of electrons with the gas;
    \textitem{3}~alignment of elongated grains on the magnetic field;
    \textitem{4}~grain rotation (relevant for the smallest sizes).
  \subsubsectbox{Dust observables}%
    A grain mixture undergoing a given set of physical conditions will 
    exhibit the following observables (represented on \reffig{fig:dustobs}):
    \textitem{1.~Emission}~the emission of a thermal continuum (\hIR\ to mm), 
      and molecular and solid state features (Mid-\hIR; \hMIR);
      a possible microwave emission (cm);
      a possible luminescence (visible);
      the possible polarization of this emission (\hIR\ to mm);
    \textitem{2.~Absorption}~the absorption of the light from a background 
      source by a continuum,
      as well as molecular and solid state features, including diffuse
      interstellar bands and ices (X-ray to \hMIR); 
      the possible polarization of this absorption (\hUV\ to visible);
    \textitem{3.~Scattering}~the scattering of the light from a bright source 
      in our direction, and its polarization (X-ray to Near-\hIR; \hNIR).
    \textitem{4.~Depletions}~some elemental depletion patterns.
  \end{textbox}

  \subsection{The Invaluable Laboratories of Nearby Galaxies}
  \label{sec:introgal}

Most of our knowledge of the dust properties comes from studies of the Milky 
Way (\hMW).
However, as it will be demonstrated in this paper, an increasing number of 
nearby galaxy studies provide unique discriminating constraints on 
fundamental dust processes.
Indeed, nearby galaxies harbor a wider diversity of environmental conditions
(metallicity, star formation activity, \etc) than what can be found in the 
\hMW.
In particular, they allow us to observe dust in extreme conditions.
Second, they constitute a necessary intermediate step towards understanding 
distant galaxies, as they are spatially resolved and have a better wavelength
coverage.
Finally, the interpretation can sometimes be more difficult in the \hMW, 
as we see the projected material of its entire disk.
On the contrary, high latitude observations of face-on galaxies can provide
cleaner sightlines.

We do not have a precise definition of \citext{nearby} galaxies.
They constitute a category expanding with the angular resolution of \hIR\ 
observatories.
In this review, we consider they are objects closer than $\simeq100$~Mpc.

  \section{THE MACROSCOPIC DISTRIBUTION OF DUST IN GALAXIES}

    \subsection{The Observational Point of View}

      \subsubsection{The Dust Distribution in Disk Galaxies}
 
Dust biases our understanding of galactic structure, as it 
affects our derivation of luminosity profiles.
The level of dust attenuation can indeed strongly vary depending on the sightline 
and from one galaxy to another \citep[\eg][]{calzetti01,pierini04,battisti16}. 
Understanding how dust is distributed in galaxies is a 
necessary step to correct for these attenuation effects.
\begin{marginnote}
  \entry{Scale-length/height}{the intensity, at radius/azimuth, $r$, can be 
         written as:
         $I(r)\propto\exp(-r/l)$, where $l$ is the scale-length/height.}
\end{marginnote}
    
\paragraph{MIR dust scale-length}
\hiso\ observations helped assess the \hMIR\ extent of nearby spirals
(especially at 7 and $15\mmic$).
They showed that the \hMIR\ and optical emission have very similar morphologies 
and concentration indices \citep{boselli03}.
However, the \hMIR\ scale-length tends to be systematically smaller than the 
optical one \citep{malhotra96} and the arm/inter-arm contrast is
larger in the \hMIR\ than in the optical \citep{vogler05}. 
From a large range of morphologies, \citet{roussel01} 
also found that if the IR-to-optical size ratio does not seem to be affected by 
the presence of a bar, 
this ratio is, on the contrary, particularly reduced in \hi-deficient 
galaxies or early-type galaxies 
\citep[see also][]{bendo02}. 
Numerous analysis have shown that the \hMIR\ disk has a similar scale-length 
to that of $^{12}$CO(1$-$0), \ha\ or the radio continuum 
\citep{sauvage96,walsh02,vogler05}.
The relation with \ha, in particular, indicates that the \hMIR\ scale-length of 
a galaxy is determined by its star-forming activity. 
Indeed, the \hMIR\ emission is enhanced in star forming regions.
With \hspitz, the improved sensitivity and 
spatial resolution opened a new frame to model the distribution of dust inside 
galaxies and in particular to derive 
radial profiles. 
\hspitz\ confirmed that the concentration index varies with wavelength. 
Studying radial profiles of 75 nearby galaxies, 
\citet{munoz-mateos09a} found that the concentration indices 
drop in the \hUIB s (5.8 and $8.0\mmic$; \refsec{sec:PAH}). 
The \hMIR\ observations at longer wavelengths 
rather indicate a large variety of behaviours, including galaxies with very 
intense nuclear, circumnuclear or outer ring emission, thus larger
concentration indices 
(\eg\ \ngc{1291}, \ngc{1512}, \ngc{1097},
\ngc{3351}).
\begin{marginnote}
  \entry{ISO}{Infrared Space Observatory 
              ($\lambda\simeq5-210\mmic$; $1995-1998$)}
  \entry{Concentration index}{\textcolor{black}{ratio between radii 
         along the major axis encompassing 75$\,\%$ and $25\,\%$ of 
         the total flux of a galaxy.}}
  \entry{Spitzer}{space telescope ($\lambda\simeq3-160\mmic$; $2003-2009$).}
  \entry{UIB}{\hypertarget{UIB}{Unidentified Infrared Bands 
              (\cf~\refsec{sec:PAH}).
              Prominent \hMIR\ features (\reffig{fig:dustobs}),
              attributed to carbonaceous grains.}}
\end{marginnote}

\paragraph{FIR/submm dust scale-length}
From \hiso\ observations, it quickly became clear that the 
disk scale-length of the \hIR\ emission increases with 
wavelength \citep{hippelein03}, with a Far-\hIR\ (\hFIR) scale-length larger 
than the optical scale-length 
\citep{tuffs96,alton98,haas98,davies99,trewhella00}.
This \hFIR\ colour gradient observed in the disk suggests that part of the 
\hFIR\ emission arises from grains heated by the radially decreasing diffuse 
InterStellar Radiation Field (\hISRF).
In the edge-on spiral galaxy \ngc{891}, 
\citet{popescu03} showed that large amounts of \hFIR\ emission was associated 
to the extended \hi\ disk, raising the question of whether grains are 
transported from the inner/optical disk, transferred via 
interactions or more resistant to destruction by shocks in the outer disk. 
\hspitz, \hhersc, as well as plane or ground-based \hsubmm\ instruments have 
since then revolutionised our vision of the \hFIR/\hsubmm\
emission and particularly confirm the detection of large cold extended disks 
\citep{block94,stevens05,hinz12}.
In \M{51}, the scale-length of the $850\mmic$ underlying exponential disk was 
estimated as $\simeq5.5$~kpc \citep{meijerink05}. 
More statistically, \citet{munoz-mateos09} and \citet{hunt15} examined the 
exponential dust profiles of $60-70$ nearby galaxies. 
They showed that the \hFIR\ scale-length does not vary strongly with galaxy 
type and is on average $\simeq10\,\%$ larger than stellar scale-lengths.
\begin{marginnote}
  \entry{Herschel}{space observatory 
                   ($\lambda\simeq55-672\mmic$; $2009-2013$).}
\end{marginnote}

      \paragraph{Scale-height Studies}
      \label{sec:scaleheight}

In the Galaxy, a scale-height of the order of 100~pc has been estimated using 
\hiras\ observations at $100\mmic$,
with a vertical distribution that correlates with the distribution of the 
\hi\ gas \citep[][]{boulanger88}.
\citet{davies97} extended the analysis to the colder dust phase using \hcobe\
data and found a more diffuse 20~K dust component, with a cold dust 
scale-height of about $\simeq500$~pc \citep{davies97}. 
Outside our \hMW, edge-on galaxies are ideal objects to constrain this 
parameter.
\citet{xilouris99} studied the stellar and dust disks in five of these 
edge-on late-type spirals and found that their mean optical-to-dust 
scale-height ratio was $\simeq1.8$. 
This ratio is often used as an \apriori\ assumption for disk models 
\citep{tempel10}. 
Quantifying the dust scale-height is undoubtedly more difficult for face-on 
galaxies. 
\citet{padoan01} proposed a new method to measure the average scale-height in 
face-on disk galaxies. 
From the \hi\ data in the Large Magellanic Cloud (\hLMC), they interpreted the 
break in the power-law shape of the spectral correlation function as 
the boundary of the gas mass distribution and velocity field. 
The method has been since then applied to derive scale-height 
estimates of the warm and cold dust. 
In the \hLMC, \citet{block10} found that the break in the power spectrum is 
occurring on scales of $\simeq100-200$~pc in the \hFIR. 
\citet{combes12} found similar scale-heights in \M{33}. 
The technique is limited by the resolution of the observations: artificial breaks 
can appear when the scale-height is too close to the resolution scale.
Finally, radiative transfer codes are robust tools to model the absorption and 
re-emission by dust and derive structural parameters (\cf~\refsec{sec:RT}).
The model of \citet{baes03}, for instance, has been used to 
derive scale-lengths and scale-heights in edge-on galaxies 
\citep{de-looze12,de-geyter15,viaene15}. 
The scale-heights derived in these studies typically range from $\simeq100$ to 
200~pc.

      \subsubsection{The Dust Distribution in Irregular Galaxies}
      \label{sec:Irr}
      
Irregular galaxies can contain large amounts of atomic gas that typically 
extend to twice their Holmberg radius \citep[\eg][]{huchtmeier81}.
They are also rich in dust, with very similar optical and \hMIR\ scale lengths 
\citep{hunter06}. 
The dust emission in irregular galaxies is clumpy and warm, with a hot dust 
and \hUIB\ emission mostly observed towards bright \hii\ regions. 
This suggests that massive stars are a major source of heating 
in these environments \citep[\eg][]{hunter06}.

      \subsubsection{The Dust Distribution in Dwarf Galaxies}
      \label{sec:dwarf}

One of the main characteristics of dwarf galaxies is their low metallicity
(\hmet).
As we will see in \refsec{sec:DG}, the dust-to-gas ratio scales roughly 
with \hmet.
The \hISM\ in these objects is less dusty and thus, more transparent.
Similarly to irregular galaxies, massive stars are a major source of heating in
these objects \citep[\eg][]{walter07}, and they are permeated by SuperNova 
(\hSN)-triggered 
shock waves \citep[\eg][]{oey96}.
Finally, these galaxies exhibit large \hi\ envelopes.
The \hIR\ emitting region can correspond to only $\simeq20-30\,\%$ of the total
mass of the system \citep[\eg][]{walter07}.

      \subsubsection{The Dust Distribution in Elliptical Galaxies}
      \label{sec:ell}

Elliptical galaxies possess very little dust: the average 
dust-to-stellar mass ratio is $\simeq50$ times lower than that of 
spiral galaxies \citep{smith12}. 
Dust-lanes are, however, commonly detected in elliptical galaxies 
\citep{sadler85b}. 
\citet{jura87} for instance found that half of nearby ellipticals are detected 
at \hiras\ wavelengths.
\citep{smith12} found that the elliptical galaxies detected at $250\mmic$ tend 
to have higher X-ray luminosities. 
Their dust may be heated in part by electron collisions \citep{goudfrooij95}.
\begin{marginnote}
  \entry{IRAS}{InfraRed Astronomical Satellite ($\lambda=12-100\mmic$; 1983).}                                    
\end{marginnote}

      \subsubsection{The Dust Distribution in Galactic Superwinds}
      \label{sec:wind}
      
Dust at high latitudes or in galactic haloes can be explained as resulting 
from various mechanisms, among which stellar feedback, transport via 
cosmic-ray driven winds or radiation pressure on the grains 
\citep{bocchio16}. 
The latter mechanism could also partly contribute to drive the galactic 
superwinds in star-forming galaxies even if several studies have shown that it 
is insufficient to be the only mechanism \citep{hopkins12,contursi13}.
\citet{contursi13} showed that, in the outflow of \M{82}, dust is slower than 
the ionized and molecular gas, indicating that dust grains are kinematically 
decoupled from the gas in the superwind. 
Most of this dust is not fast enough to escape and may fall back into the 
galaxy disk.

      \subsubsection{Dust Heating Sources Probed With Infrared Colours}
      \label{sec:colors}

Dust emits in the \hIR--\hsubmm.
In this regime, the ratio of two fluxes (or {\it colour}) provides 
information on the grains, the same way optical colours 
provide information on the stars.
These colours are widely used to understand the sources of heating of 
\hISM\ dust.
From \hiras\ observations, \citet{lonsdale-persson87} were among the first ones
to use the correlation of the \hLnu{60}/\hLnu{100} colour temperature with 
tracers of the old stars, to study their heating contribution.
With \hspitz\ observations, global \hLnu{8}/\hLnu{160} ratios 
were then used to probe the origin of the \hUIB s (\cf~\refsec{sec:PAH}) and 
to show that their emission correlates surprisingly well with that of the 
diffuse, cold dust \citep[\eg][]{bendo08}.
Resolved observations later showed that enhancements in the
\hLnu{8}/\hLnu{160} ratio are spatially offset relative to 
the star forming regions, suggesting that the \hUIB s could be partly excited 
by photons leaking out of star forming regions \citep{jones15} by up to 
$30-40\,\%$ \citep{crocker13}.
\begin{marginnote}
  \entry{Monochromatic luminosity}{\hypertarget{Lnu}{we note $\Lnu{\lambda}$, 
                                   the monochromatic luminosity 
                                   ($L_\odot/{\rm Hz}$),
                                   at wavelength $\lambda$ (\mic).}}
\end{marginnote}

With the arrival of \hhersc, the study was pushed towards longer wavelengths, 
tracing dust at lower temperatures.
\hhersc\ ratios have been extensively correlated with both
stellar surface brightnesses and star formation rate tracers. 
These analyses usually find a strong correlation of the \hLnu{250}/\hLnu{350} 
and \hLnu{350}/\hLnu{500} ratios with the local stellar mass, showing the 
importance of the lower-mass stellar populations as a heating source of the 
coldest dust population \citep{bendo10,boquien11}. 
By correlating the \hsubmm\ ratios with a linear combination of 
tracers of the total stellar population ($1.6\mmic$) and of the star forming 
regions (\ha), \citet{bendo12b} could segregate the two different 
\citext{heating sources} and found that $\simeq60-90\,\%$ of the heating of 
cold dust is assured by lower-mass stars, in disk galaxies.
Ratios at shorter wavelength, such as \hLnu{70}/\hLnu{160} or 
\hLnu{160}/\hLnu{250}, are 
less correlated with radius and more strongly correlated with the Star 
Formation Rate (\hSFR).
\citet{boselli10b} and \citet{boquien11} found similar results, showing in 
particular that the warm dust temperature as measured by the 
\hLnu{60}/\hLnu{100} ratio seems to increase with the birthrate parameter, $b$, 
whereas the cold dust temperature, (measured by the \hLnu{350}/\hLnu{500}
ratio) seems to be anti-correlated with $b$.
However, the old stellar population probably continues to also play a role in 
the heating of the warm dust, with a contribution that seems to correlate with 
the galaxy evolutionary stage \citep[for instance significant
global contribution of the bulge stars in early-type galaxies like 
\M{81};][]{bendo12b}.
On the other hand, \citet{remy-ruyer13} showed a trend of colour
temperature with metallicity, suggesting that low-\hmet\ systems are on average
hotter \citep[see also][]{melisse94b}.
This is conjectured to be due to the enhanced contribution of 
young star heating at low-\hmet.
\begin{marginnote}
  \entry{Birth rate parameter}{current \hSFR\ divided by the mean \hSFR\ over 
                               the lifetime of the galaxy.}
\end{marginnote}

New results from radiative transfer models (\cf~\refsec{sec:RT}) are now 
quantifying better the respective contributions of the different stellar 
populations to dust heating. 
In \M{31}, \citet{viaene17} showed that $90\,\%$ of the dust could be
heated by the lower-mass stellar populations (see their Fig.~8). 
Further detailed analysis would be necessary to quantify more robustly the 
contribution as a function of the galaxy type.

    \subsection{The Radiative Transfer Approach}
    \label{sec:RT}
    
\begin{textbox}[h]
  \sectboxns{WHAT ARE THE SPATIAL SCALES RELEVANT TO DUST HEATING?}%
The dust physical conditions vary on spatial scales of the order of the mean 
free path of a \hV\ photon:
\begin{equation}
  l_{V} = \left[
  \underbrace{\overbrace{Y_\sms{dust}}^\sms{dust-to-H mass ratio}
  \times \overbrace{m_\sms{H}}^\sms{mass of an H}
  \times \overbrace{n_\sms{H}}^\sms{density}}_\sms{dust mass per unit ISM 
                                                   volume}
          \times\underbrace{\kappa(V)}_\sms{dust opacity}\right]^{-1}
        \simeq \frac{600{\;\rm pc\,cm^{-3}}}{n_\sms{H}}.
  \label{eq:meanfreepath}
\end{equation}
For a diffuse region ($n_\sms{H}\simeq 10\;\rm cm^{-3}$), 
$l_\sms{V}\simeq60$~pc, while for a dense region 
($n_\sms{H}\simeq 10^4\;\rm cm^{-3}$),
$l_\sms{V}\simeq0.06$~pc.
Thus, to resolve dust temperature variations, in an edge-on cloud, we would 
need to resolve structures of $\simeq0.06$~pc, which translates into
$\simeq0.2^{\prime\prime}$ for the \hLMC\ (typical \hALMA\ \&\ \hjwst\ 
resolution) and $\simeq0.01^{\prime\prime}$ in \M{31} (currently inaccessible).
For a face-on cloud, there will be mixing along the line of sight, in any 
case.
\end{textbox}

      \subsubsection{Radiative Transfer Models}
      \label{sec:RTmod}
      
The most rigorous way to understand the effects of dust extinction and emission 
on \hUV-to-mm observations of galaxies is to model the Radiative Transfer 
(\hRT) of the starlight through their dusty \hISM, in a realistic 3D geometry.
Several teams have developed such codes for disk galaxies
\citep[\eg][]{baes03,bianchi08,popescu11}.
These codes solve the \hRT\ equation, accounting for multiple 
anisotropic scattering, absorption, and dust and stellar emission.
This type of computation is numerically intensive.
Most models implement a Monte Carlo method, with various refinements and
heavy parallelization.

      \subsubsection{Application to Galaxies}
      \label{sec:RTobs}

\begin{figure}[h]
 \includegraphics[width=1.23\textwidth]{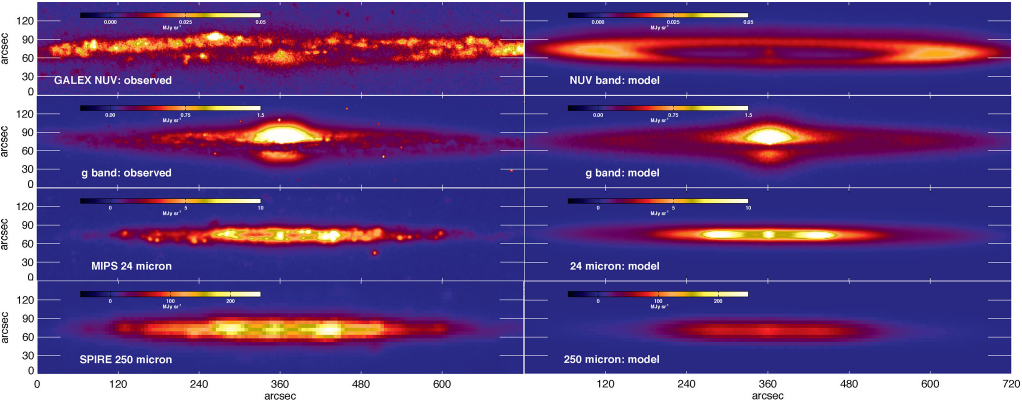}
  \caption{{\sl \hRT\ model of \ngc{4565} \citep{de-looze12}}.
           The observations (left column) are compared to the model
           (right column).}
  \label{fig:RT}
\end{figure}
Applying 3D \hRT\ models to reproduce the spatial flux distribution of 
galaxies, in all wavebands, is not straightforward.
Indeed, the observations provide only 2D projected constraints.
This is why most studies favor edge-on galaxies, as
the images of such objects provide constraints on both
the radial and azimuthal distributions, assuming axisymmetry (\reffig{fig:RT}).
Several studies have modelled the effect of extinction on the optical data of 
disk galaxies using such codes \citep[\eg][]{xilouris99,alton04,bianchi07}.
They were able to answer the recurring question about the optical thickness of 
disk galaxies \citep{disney89}.
In particular, \citet{xilouris99} found that the face-on optical depth of 
typical spiral galaxies is less than one in all optical bands.
These studies also provide a more comprehensive answer to the nature of the 
dust heating sources and on the scale-length and scale-height  
(\cf~\refsecs{sec:colors}{sec:scaleheight}).
Finally, these models account for the energy balance between the escaping
\hUV-visible light and the re-emitted \hIR-\hsubmm\ radiation.
Interestingly enough, several studies report a deficit of modelled
\hFIR\ emission by a factor $\simeq3-4$, compared to the observations
\citep{alton00,alton04,dasyra05,de-looze12,de-looze12b}.
This discrepancy is thought to reside in a lack of details in modelling the 
geometry.
In particular, the presence of young stars, deeply embedded in molecular 
clouds, could compensate this deficit without significantly altering the 
extinction \citep[\eg][]{baes10}.

    \subsection{Phenomenological SED Modelling}

      \subsubsection{The Mixing of Physical Conditions}
      \label{sec:mixT}

Radiative transfer is impractical in most cases, as the geometry of the source
is often poorly known.
In addition, radiative transfer models of whole galaxies
do not resolve spatial scales small enough \refeqp{eq:meanfreepath}.
Hence, we are usually compelled to make approximations about the complex
topology of the studied object.

\begin{figure}[h]
  \includegraphics[width=1.23\textwidth]{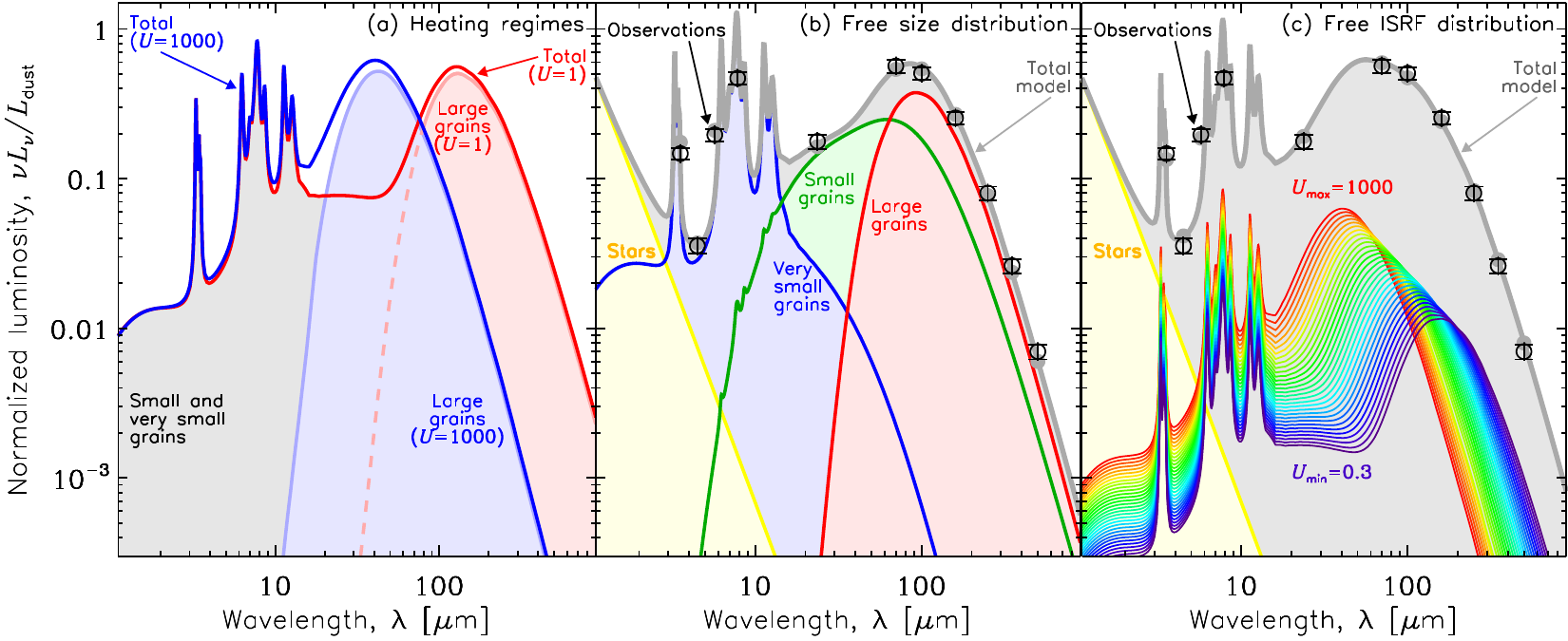}
  \caption{{\sl Phenomenological SED modelling.}
           \textitem{a}~Dust emission model of the \hMW\ 
           \citepalias{jones17}, heated by the \hISRF\ of the diffuse
           \hISM\ ($U=1$; red) and higher ($U=1000$; blue).
           It demonstrates that the shape of the \hSED\ of small, 
           out-of-equilibrium grains (grey), is independent of \hU. 
           On the contrary, the shape of the \hSED\ of large grains,
           which are at equilibrium with the \hISRF, shifts to shorter 
           wavelengths when \hU\ increases, as their equilibrium temperature
           increases.
           \textitem{b}~Simulated broadband observations (black),
           fitted by varying the mass fractions of 
           {\it very small} ($0.35\le r<1.5\,$nm; blue), 
           {\it small} ($1.5\le r<20\,$nm; green)
           and {\it large} ($0.02\le r<0.3\mmic$; red) grains,
           with a single $U=8$ (fitted).           
           \textitem{c}~Same simulated observations as in panel~(b), 
           alternatively fitted 
           with a starlight intensity distribution ($\propto U^{-\alpha}$; 
           rainbow curves).}
  \label{fig:fitSED}
\end{figure}

\paragraph{The Isothermal approximation}
\label{sec:MBB}
To first order, one can ignore the variations of the physical conditions.
The Modified Black Body (\hMBB), the most widely used 
approximation, assumes
that the \hIR\ emission comes from identical grains, at a single temperature,
$T_\sms{dust}$, with a power-law, wavelength-dependent, dust mass absorption 
coefficient, or \citext{opacity}:
\begin{equation}
  \underbrace{L_\nu(\lambda)}_\sms{monochromatic luminosity}
    = \underbrace{M_\sms{dust}}_\sms{dust mass}
      \times 
      \underbrace{\kappa(\lambda_0).
                 \left(\lambda_0/\lambda\right)^\beta}_\sms{parametric opacity}
      \times
      \underbrace{4\pi B_\nu(\lambda,T_\sms{dust})}_\sms{black body}.
  \label{eq:MBB}
\end{equation}
In principle, a fit of this model, varying $M_\sms{dust}$, $T_\sms{dust}$ and 
\hbeta, provides constraints both on the grain physical conditions (through 
$T_\sms{dust}$), and on their composition (through \hbeta).
Indeed, different materials can have different \hbeta.
An inverse $T_\sms{dust}-\beta$ relation is also observed on some 
laboratory analogs \citep[\eg][]{mennella98,boudet05}. 
However, a gradient of temperature tends to broaden the \hSED.
The inherent mixing of physical conditions is thus enough to bias these 
estimates \citep[\eg][]{juvela12,hunt15}.
In addition, the contribution from out-of-equilibrium grains can be
non negligible at $\lambda\lesssim70\mmic$ (\refpanel{fig:fitSED}{a}).
In that sense, the \hMBB\ derived parameters are ambiguous effective 
quantities ($T_\sms{eff}$, $\beta_\sms{eff}$), which can be reliably 
interpreted in terms of intrinsic grain properties only:
\textitem{1}~in diffuse regions, where the \hISRF\ is 
expected to be roughly uniform, or in cold cores;
and \textitem{2}~provided that the fit is constrained at long enough 
wavelengths ($\lambda\gtrsim100\mmic$).
\begin{marginnote}
  \entry{ISRF intensity}{\hypertarget{U}{\hU\ 
                         is the \hISRF\ intensity, integrated in 
                         $[0.0912,8]\mmic$.
                         It is normalized so that $U=1$ in the solar 
                         neighborhood.}}
\end{marginnote}

Alternatively, one can also fit an observed \hSED, with a full dust mixture 
(such as in \refpanel{fig:fitSED}{a}), varying the \hISRF\ intensity, \hU, and 
the mass of each sub-components (\refpanel{fig:fitSED}{b}).
With such an approach, the \hMIR\ wavelengths can be interpreted in terms of 
size distribution variations, provided that the \hISRF\ is roughly uniform.

\begin{textbox}[h]
  \sectbox{DUST HEATING REGIMES: EQUILIBRIUM OR STOCHASTIC?}
  \subsubsectbox{Thermal Equilibrium}
    The enthalpy, $H$, of grains with large enough radii ($r\gtrsim0.02\mmic$)
    is, in most cases, significantly higher than the mean energy of the 
    incident photons they absorb, $\langle h\nu\rangle\ll H$.
    Therefore, a single photon event does not notably modify their temperature.
    They are at equilibrium with the radiation field.
    Their spectrum is proportional to a Planck function times a 
    wavelength-dependent opacity (\refpanel{fig:fitSED}{a}).
  \subsubsectbox{Stochastic Heating}
    On the opposite, for small grains ($r\lesssim0.02\mmic$), 
    $\langle h\nu\rangle\gtrsim H$.
    A single photon event will cause temperature spikes at a few hundred K
    (depending on its size), followed by a significant cooling before the next 
    absorption \citep{draine85}.
    These temperature fluctuations result in a broad spectrum, extending down 
    to the MIR (\refpanel{fig:fitSED}{a}).
    Such grains are out of thermal equilibrium.
\end{textbox}

\paragraph{Distributions of starlight intensities}
\label{sec:Udist}
It is possible to empirically account for the mixing of physical conditions, by
parameterizing the \hISRF\ intensity distribution.
One of the most useful prescriptions is given by \citet{dale01}.
It assumes that the dust mass in different bins of \hU\ follows a power-law:
$dM_\sms{dust}\propto U^{-\alpha}dU$ for $U_\sms{min}<U<U_\sms{max}$.
This way, varying the parameters $U_\sms{min}$, $U_\sms{max}$ and $\alpha$,
an observed \hSED\ can be fit with a combination of physical conditions.
This is demonstrated in \refpanel{fig:fitSED}{c}.
One of the limitations of this approach is that it ignores the variations of 
the grain properties with environments.
As will be discussed in \refsec{sec:localevol}, the carriers of the \hUIB s 
are usually destroyed in regions of high \hU.
The \hFIR\ emissivity will also change due to mantle processing and 
evaporation (\cf~\refsec{sec:FIRevol}).
In addition, there is a degeneracy between the \hISRF\ distribution and the
fraction of small grains.
This is demonstrated in panels~b and c of \reffig{fig:fitSED}.
The same \hSED\ is fitted either with an isothermal model, accounting for the 
\hMIR\ fluxes by raising the fraction of small grains (b); or by adding hotter 
regions (c).
Fortunately, the dust mass is dominated by the coldest large grains, and can
thus be reasonably derived with this type of model.
This type of model is superior to \hMBB, as the latter underestimates the 
mass by $\simeq30\,\%$ in rather diffuse regions, and down to a factor of 
$\simeq3$, where there is a lot of mixing 
\citep[\eg][]{galliano11,galametz12,remy-ruyer15}.

More complex parameterizations of the \hISRF\ distribution are possible
\citepalias[\eg][]{draine07}.
It is also possible to build multi-component models, where the phase 
composition is linked to the star formation history \citep[\eg][]{da-cunha08}.

\paragraph{A Matriochka effect}
A consequence of the mixing of physical conditions is that the derived dust 
mass depends on the spatial resolution.
Indeed, cold regions have a weak luminosity, but can be massive.
When a large region is integrated, the cold components can be hidden, while 
they can be accounted for in smaller resolution elements.
In practice, the derived dust mass is always higher (up to 
$\simeq50\,\%$), when estimated at high spatial resolution, than on integrated 
fluxes \citep{galliano11,galametz12,roman-duval14}.
The estimate at high resolution is thought to be more accurate.
This result depends on the environment, on the maximum resolution and on the
parameterization of the \hISRF\ distribution.
It is not always seen \citep[\eg][]{aniano12}.

    \subsubsection{Fitting Methodologies}
    \label{sec:SEDfit}

\hSED\ fitting is a technical matter.
First, the contribution to the broadband fluxes of several non-dust-related
processes, need to be subtracted, or at least accounted for in the 
uncertainties (\reffig{fig:dustobs}):
\textitem{1}~the free-free and synchrotron radio continuum;
\textitem{2}~ionic, \hPDR\ and molecular lines;
\textitem{3}~foreground Galactic and zodiacal emission;
\textitem{4}~background galaxies.
Second, finding the best model parameters is not straightforward.
The least-squares method is the most widely used approach to fit 
an \hSED\ model to \hIR\ observations.
However, \citet{shetty09} have shown that this approach leads to a 
noise-induced false anticorrelation between $T_\sms{dust}$ and \hbeta, in the 
case of a \hMBB.
It makes the determination of the possible real $T_\sms{dust}-\beta$ relation 
(\refsec{sec:MBB}) difficult.
\citeauthor{galliano18} ({\it submitted}; hereafter \citetalias{galliano18})
showed that similar noise-induced correlations are found 
between most parameters, when using full dust models with an \hISRF\ 
distribution.

Bayesian inference is becoming increasingly popular with the development of 
powerful computers.
Several \hSED\ models implement it to provide rigorous uncertainty estimates
of the derived parameters.
However, \citetalias{galliano18} has shown that this approach is not sufficient 
to solve the noise-induced correlations.
In fact, \citet{kelly12} have demonstrated, in the case of a \hMBB, that 
it is necessary to use the more complex \citext{hierarchical Bayesian} 
inference to achieve this goal.
\citetalias{galliano18} has extended this technique to full dust models with
\hISRF\ distributions, and has shown that it leads to significant improvements 
in the recovery of the dust parameters and of their intrinsic correlations.

    \subsection{The Grains in Relation with the Gas and the Stars:
                Diagnostic Tools}
    
      \subsubsection{Dust-Related Star Formation Rate Indicators}
      \label{sec:SFR}

The \hSFR\ is a crucial quantity for galaxy evolution studies.
Several dust-related \hSFR\ tracers have been empirically calibrated using 
observations of resolved star forming regions in nearby galaxies.
They rely on the fact that young stars are extremely luminous and are 
enshrouded with dust. 
If the clouds are optically thick and if their covering factor is unity, the
OB star luminosity is: $L_\sms{OB}\simeq L_\sms{TIR}$.
Contrary to a common misconception, this is independent of dust properties.
Assuming a typical Initial Mass Function (\hIMF), burst age and metallicity, 
$L_\sms{OB}$ can be converted to: 
${\rm SFR}/(M_\odot/{\rm yr})\simeq 10^{-10}\times L_\sms{TIR}/L_\odot$.
The contribution of old stars to $L_\sms{TIR}$ is negligible for high enough 
\hSFR s.
Alternatively, monochromatic \hSFR\ indicators have been proposed.
\citet{calzetti07} and \citet{li10} found that the 24 and $70\mmic$
monochromatic luminosities were good local \hSFR\ indicators (on spatial scales 
of $\simeq 0.5-1$~kpc):
${\rm SFR}/(M_\odot/{\rm yr})\simeq 2611\times [\hmLnu{24}/(L_\odot/{\rm Hz})]^{0.885}$
and 
${\rm SFR}/(M_\odot/{\rm yr})\simeq 1547\times \hmLnu{70}/(L_\odot/{\rm Hz})$.
Otherwise, \citet{peeters04} found that, although the $6.2\mmic$ \hUIB\ 
correlates with \hSFR, it is probably a better B star tracer.
Moreover, the \hUIB\ strength is strongly metallicity dependent 
(\refsec{sec:formPAH}).
Finally, several composite indicators have been calibrated \citep{hao11}.
By combining Far-\hUV\ (\hFUV) or \ha\ measurements with the $24\mmic$ or 
\hTIR\ indicators, they account for the fact that star forming regions are not 
completely opaque.
\begin{marginnote}
  \entry{TIR}{\hypertarget{TIR}{total infrared ($\lambda=3-1000\mmic$).
                                $L_\sms{TIR}$ is the integrated power in this
                                range.}}
\end{marginnote}

      \subsubsection{Estimating the Total Gas Mass}  
      \label{sec:gas}
      
The complexity of the \hISM\ phase mixing makes the total gas mass 
difficult to estimate.
The ionized, neutral atomic and molecular phases, require each one an 
independent tracer.
And even when these tracers are available, some notable biases question the 
reliability of these estimates:
\textitem{1}~the \hiline\ emission can be saturated at high optical depth,
leading to a possible underestimate of \N{\hi} by a factor up to 
$\simeq2$ \citep[\eg\ in the local \hISM;][]{fukui15};
\textitem{2}~the \COio, used to estimate \N{\hmol}, can be biased by the
selective photodissociation of CO in translucent \hmol\ clouds, leading to 
a possible underestimate of \N{\hmol} by a factor up to $\simeq100$
\citep[\eg\ in low-\hmet\ systems;][]{madden97}.
There is thus a reservoir of \citext{dark gas}, unaccounted for by these 
tracers.
\hUV\ \hi\ and \hmol\ absorption lines are less biased, but are difficult to 
observe.

Whether observed in extinction or in emission, dust has been used as an 
indirect gas tracer for several decades \citep{hildebrand83,devereux90}.
Neglecting the ionized gas, and assuming that the \hi\ surface density, 
\Surf{\hi}, 
has been corrected for optical thickness, the dust surface density can be 
expressed as:
$\mSurf{dust}=\DG\times(\mSurf{\hi}+\alpha_\sms{CO}I_\sms{CO})$,
where $\alpha_\sms{CO}$ is the conversion factor between the CO intensity, 
$I_\sms{CO}$, and the \hmol\ surface density.
\citet{israel97} pioneered in deriving $\alpha_\sms{CO}$
in nearby galaxies, using the \hFIR\ dust emission.
\citet{leroy11} designed a method to solve both for \hDG\ and 
$\alpha_\sms{CO}$, assuming a homogenous grain constitution and a constant
\hDG\ in each one of the objects they studied.
They confirmed that low-\hmet\ galaxies have larger $\alpha_\sms{CO}$ than 
the \hMW\ (up to a factor of $\simeq20$).
This well-established fact is believed to originate in the enhanced 
photodissociation of CO in an \hISM\ less dusty, thus less opaque 
(\cf~\refsec{sec:DG}), while \hmol\ remains self-shielded.
Similar studies have been done in extinction 
\citep[\eg\ in the \hLMC;][]{dobashi08}.
\begin{marginnote}
  \entry{Dust-to-gas mass ratio}{\hypertarget{DG}{noted \hDG, defined as
                                 $\hmDG=\mSurf{dust}/\mSurf{gas}$,
                                 where \Surf{gas} is the total gas mass
                                 surface density
                                 (\hii, \hi\ and \hmol, including He).}}
\end{marginnote}

      \subsubsection{Constraints for Photodissociation Models} 
      \label{sec:PDR}

In \hPDR s, the gas is primarily heated by photoelectrons ejected from the 
grains \citep{draine78}.
The \hPE\ heating is more efficient for the smallest sizes, in particular, for 
the carriers of the \hUIB s \citep[\eg][]{weingartner01b}.
This heating efficiency therefore depends on the dust properties, and thus on 
the environment.
Assuming that \ciiline\ is the main gas coolant, the
\hPE\ efficiency, $\epsilon_\sms{PE}$, can be approximated by the gas-to-dust 
cooling ratio: $\epsilon_\sms{PE}\simeq L_\sms{\cii}/L_\sms{TIR}$.
Detailed studies usually add other lines to the gas cooling rate, like \oiline,
to have a more complete estimate 
\citep[\eg][]{cormier15}.
Overall, \citet{smith17} found that 
$0.1\,\%\lesssim \epsilon_\sms{PE}\lesssim 1\,\%$, with an average of 
$\langle\epsilon_\sms{PE}\rangle\simeq0.5\,\%$, in a sample of 54 nearby 
galaxies.
It appears that $\epsilon_\sms{PE}$ is lower when the dust temperature is 
higher \citep{rubin09,croxall12}.
This is not likely the result of the destruction of the \hUIB\ carriers, as
their intensity correlates the best with the \ciiline\ emission 
\citep[\eg][]{helou01}.
It is rather conjectured to be due to the saturation of grain charging 
in \hUV-bright regions.
The shape of the \hISRF\ also has a consequence on the accuracy with which 
$L_\sms{TIR}$ represents the true \hUV, \hPE-efficient flux.
Indeed, \citet{kapala17} showed that the variations of $\epsilon_\sms{PE}$ in 
\M{31} could be explained by the contribution of old stars to $L_\sms{TIR}$.
Finally, one of the most puzzling features is that $\epsilon_\sms{PE}$ is 
higher at low metallicity \citep{poglitsch95,madden97,cormier15,smith17},
while the \hUIB\ strength drops in these systems (\refsec{sec:formPAH}).
This is currently poorly understood.
However, in the extreme case of I$\,$Zw$\,$18 ($Z\simeq1/35\,Z_\odot$), 
where no \hUIB\ is detected and the \hPE\ heating is estimated to be 
negligible, the gas-cooling-to-\hTIR\ ratio is still $\simeq1\,\%$ 
\citep{lebouteiller17}.
In this instance, the gas could be heated by X-rays.

Now, studies focusing on the gas properties usually run \hPDR\ models with
build-in \hPE-efficiency \citep[\eg][]{le-petit06}.
In this case, dust parameters such as $L_\sms{TIR}$, the visual attenuation or 
the \hUIB\ strength help refine the gas modelling
\citep[\eg][]{chevance16}.

  \section{CONSTRAINTS ON THE MICROSCOPIC DUST PROPERTIES}

    \subsection{The Infrared Emission}

\begin{figure}[h]
  \includegraphics[width=1.23\textwidth]{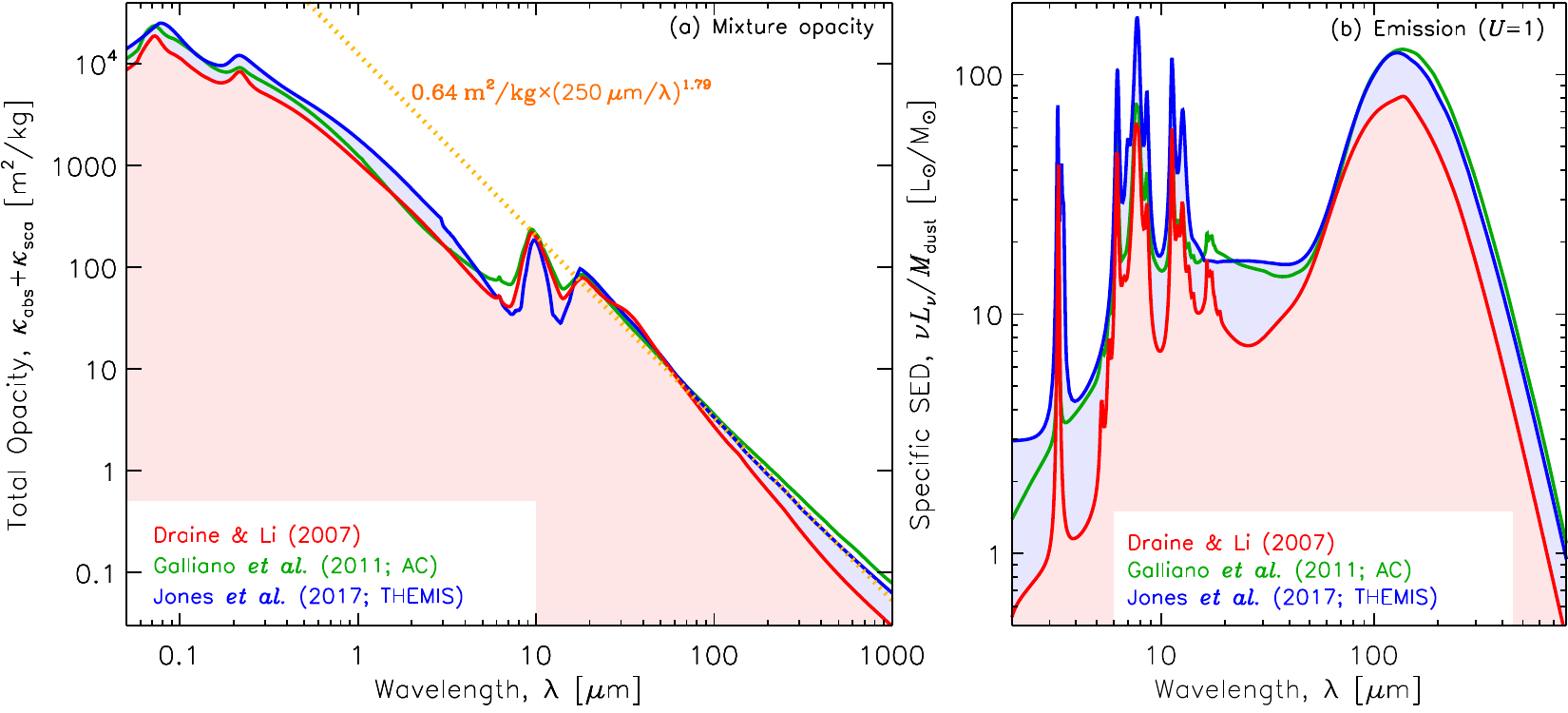}
  \caption{{\sl Emissivity of select dust models.}
           \textitem{a}~Extinction opacities. 
           Notice the different submm slopes.
           A power-law approximation to \citetalias{jones17} is shown 
           in dashed yellow ($\beta=1.79$, $\lambda_0=250\mmic$,
           $\kappa(\lambda_0)=0.64\;\rm m^2\,kg^{-1}$).
           \textitem{b}~Corresponding emissivities.}
  \label{fig:kappa}
\end{figure}

      \subsubsection{Constraints on the FIR Grain Opacity}
      \label{sec:FIR}

The dust mass and excitation derived from fits of the \hSED\ 
depend directly on the assumed grain opacity.
To first order, it can be approximated by a power-law, in the 
\hFIR/\hsubmm\ (\reffig{fig:kappa}{a}).
Most studies assume a fixed absolute opacity, $\kappa(\lambda_0)$, and explore 
the variations of the emissivity index, \hbeta.

\begin{table}[htbp]
  \caption{Free emissivity index MBB fits of nearby galaxies 
           by the \planck\ collaboration.}
  \label{tab:MBBfit}
  \begin{tabularx}{\textwidth}{l|*{4}{X}}
    \hline
    \hline
      \rowcolor{coltab1}
      & Milky Way & \M{31} & LMC & SMC \\
    \hline
      \cellcolor{coltab2} $T_\sms{eff}$ & $19.7\pm1.4$~K 
        & $18.2\pm1.0$~K & $21.0\pm1.9$~K & $22.3\pm2.3$~K \\
    \hline
      \cellcolor{coltab2} $\beta_\sms{eff}$ & $1.62\pm0.10$ 
        & $1.62\pm0.11$ & $1.48\pm0.25$ & $1.21\pm0.27$ \\
    \hline
      \cellcolor{coltab2} Reference & \citetalias{planck-collaboration14c} 
        & \citetalias{planck-collaboration15c} 
        & \citetalias{planck-collaboration11} 
        & \citetalias{planck-collaboration11} \\
    \hline
  \end{tabularx}
\end{table}  
\paragraph{Studies of the emissivity index}
\label{sec:beta}
There are numerous publications presenting \hMBB\ fits of nearby galaxies.
However, as discussed in \refsec{sec:mixT}, the derived \hbeff\ is 
degenerate with temperature mixing.
The best constraints on the intrinsic \hbeta\ are obtained in the submm regime,
where only massive amounts of very cold dust ($T\lesssim 10$~K) could bias the 
value.
\reftab{tab:MBBfit} lists \hbeff\ for several objects, 
obtained with \hplanck, with constraints up to $850\mmic$.
It appears that all the values are lower than 2, and that low-\hmet\ systems 
have a lower \hbeff\ than higher \hmet\ galaxies.
\citet{boselli12} studying a volume-limited sample with \hhersc\ (up to 
$500\mmic$) also found an average $\beta_\sms{eff}\simeq1.5$, and hinted that 
low-\hmet\ objects tends to have $\beta_\sms{eff}<1.5$.
In \M{33}, \hbeff\ derived from \hersc\ observations is around
2 in the center and decreases down to 1.3 in the outer parts 
\citep{tabatabaei14}.
On the other hand, the outer regions of \M{31} exhibit a steeper slope 
($\beta_\sms{eff}\simeq2.3$) than in its center \citep{draine14}.
This contradictory behaviour does not appear to originate in fit biases, as 
both increasing and decreasing trends of \hbeff\ with radius are 
found in the sample of \citet{hunt15}.
\begin{marginnote}
  \entry{Planck}{\hypertarget{planck}{\hFIR--cm space telescope 
                                        ($\lambda\simeq300\mmic-1\,$cm; 
                                        $2009-2013$).}}
\end{marginnote}

\paragraph{Constraints on the absolute opacity}
\label{sec:kappa0}
The absolute opacity, $\kappa(\lambda_0)$, is totally degenerate with 
$M_\sms{dust}$ \refeqp{eq:MBB}. 
The only way to constrain this quantity is to have an independent estimate of 
$M_\sms{dust}$.
For instance, \citet{galliano11} modelled the IR emission of a strip covering
1/4 of the \hLMC, using optical properties similar to 
\citet[][hereafter \citetalias{draine07}]{draine07}.
The resulting dust-to-gas mass ratio was higher than the maximum number of 
elements that could be locked-up in grains.
It led them to propose a more emissive mixture (\reffig{fig:kappa}) 
that could solve this inconsistency.
The constraint on the elemental depletions was thus used to show 
$\kappa(\lambda_0)$ should be a factor of $\simeq2-3$ higher.
Quite similarly, \citet{planck-collaboration16} modelled the all sky dust 
emission using the \citetalias{draine07} model.
However, the \Av\ estimated along the sightlines of $\gtrsim200\,000$ quasars 
was systematically lower than their dust-emission-derived \Av.
Their comparison of emission and extinction thus indicates that the Galactic 
opacity should also be a factor of $\simeq2$ higher.
Finally, in \M{31}, \citet{dalcanton15} derived a high spatial resolution 
map of \Av.
As in the Galaxy, the emission-derived \Av\ map \citep{draine14} is found to be 
a factor of $\simeq2.5$ higher.
We emphasize that, although each of these studies found evidence of local 
variations of the emissivity as a function of the density 
(\cf~\refsec{sec:FIRevol}), the overall opacity seems to be scaled up compared 
to \citetalias{draine07}.
There is growing evidence that the 
dust opacities might be, in general, more emissive than standard uncoated 
compact silicate-graphite mixtures.

      \subsubsection{The Aromatic Feature Spectrum}
      \label{sec:PAH}
    
\begin{figure}[h]
  \includegraphics[width=1.23\textwidth]{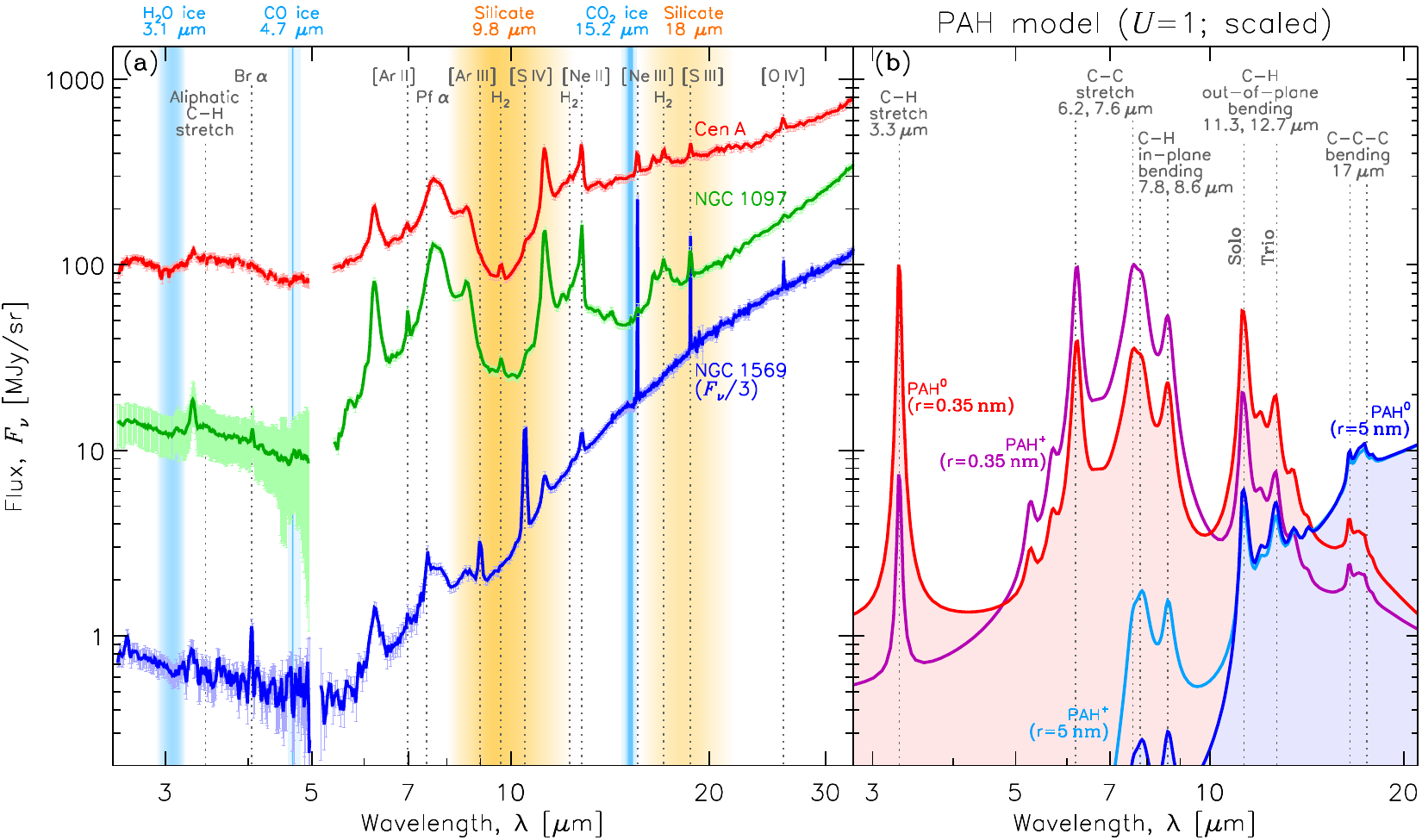}
  \caption{{\sl MIR spectra.} 
           \textitem{a}~\hakari\ ($2.5-5\mmic$) and 
           \hspitz\ ($5-30\mmic$) spectra of the central regions of
           three galaxies (T.~Roland \&\ R.~Wu, {\it priv.\ comm.}).
           \cena\ is an early-type galaxy with an AGN; it shows strong silicate
           absorption.
           \ngc{1097} is a late-type starburst with a weak AGN.
           \ngc{1569} is a blue compact dwarf, exhibiting weak AFs and 
           strong ionic lines.
           \textitem{b}~Theoretical emission of ionized and neutral PAHs 
           \citepalias{draine07}, with radii $r$, heated by the diffuse 
           Galactic ISRF ($U=1$).
           These spectra are scaled by an arbitrary factor.
           Solo, trio: one or three C--H bonds on a given aromatic cycle.}
  \label{fig:MIR}
\end{figure}
The \hMIR\ range exhibits a complex spectrum of ubiquitous emission bands, 
that were originally labeled as \hUIB s (\reffig{fig:MIR}).
These bands are found in almost every environment.
They were first detected in Galactic planetary nebulae \citep{gillett73}.
Their relative intensity appeared independent of position in reflection 
nebulae, leading to the conclusion that they were carried by very small, 
stochastically heated particles \citep{sellgren84}.
\begin{marginnote}
  \entry{Akari}{space telescope ($\lambda\simeq1.7-180\mmic$; $2006-2011$).}
\end{marginnote}

\paragraph{Candidate materials} 
It is consensual, nowadays, that the \hUIB s arise from the vibrational 
spectrum of a collection of aromatic bonds in a hydrocarbon matrix 
\citep[the weak $3.4\mmic$ feature is attributed to aliphatic bonds;][]{duley81}.
\hPAH s were proposed by \citet{leger84} and \citet{allamandola85}.
The \hUIB s likely arise from a statistical mixing of \hPAH s of different sizes 
and structure, averaging out drastic variations of the band profiles.
Most studies interpret their \hMIR\ spectra, in light of this class of 
molecules.
The main physical processes controlling the shape of the \hPAH\ spectrum are 
the following.
\begin{description}
  \item[Charge.]
    \hPAH$^+$ have brighter $6-9\mmic$
    features than \hPAH$^0$, and inversely for the other bands 
    (\refpanel{fig:MIR}{b}).
    Select band ratios, such as \hLint{7.7}/\hLint{11.3}, increase with the 
    charge.
  \item[Size distribution.] 
    Small \hPAH s fluctuate up to temperatures higher than the large ones.
    Short wavelength bands are therefore more pumped in small \hPAH s, while
    large \hPAH s emit predominantly long wavelength features 
    (\refpanel{fig:MIR}{b}).
  \item[ISRF hardness.]
    The hardness of the \hISRF\ has an effect similar to the size, as a higher 
    mean photon energy causes the grain to fluctuate up to higher temperatures
    \citep[\eg][hereafter \citetalias{galliano08b}]{galliano08b}.
  \item[Molecular structure.]
    C--H out-of-plane bending modes have different frequencies, depending
    on the number of H atom per aromatic cycle.
    The $11.3\mmic$ band corresponds to a solo H, found on straight molecular 
    edges, while the $12.7\mmic$ one
    corresponds to a trio, found on corners of the molecules 
    (\refpanel{fig:MIR}{b}).
    The solo/trio \hLint{11.3}/\hLint{12.7} ratio is thus an indicator of
    \hPAH\ compactness 
    \citep[\eg][]{hony01}.
  \item[Foreground extinction.]
    The wings of the silicate absorption feature at $9.8\mmic$ selectively 
    extinct more the 8.6 and $11.3\mmic$ band than the other features
    (\refpanel{fig:MIR}{a}).
  \item[Dehydrogenation.]
    It has a similar effect to ionization.
    However, for \hPAH s larger than $N_\sms{C}\simeq25$, 
    hydrogenation through reactions with abundant atomic H is more
    important than H loss through unimolecular dissociation
    \citep[\eg][]{hony01}.
    Thus, dehydrogenation does not have a detectable effect on the \hUIB\ 
    spectrum.
\end{description}
However, a mix of fully aromatic and/or partially hydrogenated molecules does 
not appear to give a completely satisfactory fit to the wavelength position and 
shape of the observed interstellar emission bands in the $3.3-3.5\mmic$ region 
\citep[\eg, Fig.~3 in][]{sandford13}. 
Alternatively, the \citetalias{jones17} model \hHAC\ nanoparticles could 
provide a physically-viable alternative to the interstellar \hPAH\ hypothesis 
for the carriers of the \hUIB s. 
To date, the \citetalias{jones17} interstellar dust model appears to be the 
only one that is consistent with the presence of the emission bands at 
$\simeq3.3$ and $\simeq3.4\mmic$ and the associated sub-bands at 
$\simeq3.5\mmic$. 
In this model, these bands are due to \hHAC\ nano-grains, of mixed aromatic, 
olefinic and aliphatic composition, with size-dependent optical properties 
\citep{jones13}. 
Further, and within the framework of \citetalias{jones17}, variations in the 
band ratios in 
the $3.3 - 3.5\mmic$ region can be explained by \hHAC\ compositional variations 
as the dust evolves in response to its  environment. 
That is the reason why we adopt the generic acronym \hUIB.
We refer to \hPAH s only when discussing results depending on the \hPAH\ 
hypothesis.
\begin{marginnote}
  \entry{PAH}{\hypertarget{PAH}{polycyclic aromatic hydrocarbons. 
              Molecules composed of several aromatic cycles, with 
              peripheral H atoms.}}
  \entry{\HAC}{\hypertarget{HAC}{amorphous carbon with partial hydrogenation.
               Solids containing both aromatic and aliphatic bonds,
               in proportion depending on the H-fraction.}}
               \label{pipo}
\end{marginnote}

\paragraph{MIR spectrum fitting challenges}
To study the properties of the \hUIB s, one needs to estimate the 
intensity of each band.
Most studies perform least-squares spectral decompositions, with the linear 
combination of several components (\eg\ \citealt{smith07}, hereafter 
\citetalias{smith07}; \citetalias{galliano08b}):
\textitem{1}~\hPAH s: Lorentz or Drude profiles;
\textitem{2}~gas lines: Gauss profiles;
\textitem{3}~dust continuum: several \hMBB s;
\textitem{4}~a stellar continuum.
In addition, \textitem{5}~the extinction by silicate and ices can be 
parameterized by their optical depths.
However, as can be seen on \refpanel{fig:MIR}{a}, the wings of the \hUIB s can 
be difficult to distinguish from the dust continuum.
In addition, several features are blended (\eg\ the $6-9$ or $17\mmic$ 
complexes) and may present underlying broad plateaus.
These bands can also be blended with ionic lines, at low spectral resolution
(\eg\ the $12.7\mmic$ and \neiiline).
This type of fit can thus lead to some biases and degeneracies in the following 
cases \citepalias{galliano08b}:
\textitem{1}~low signal-to-noise ratio;
\textitem{2}~low \hUIB-to-continuum ratio;
\textitem{3}~moderate extinction.
The derived band intensities are also sensitive to the adopted \hUIB\
positions and widths, and on the nature and number of \hMBB\ components for the 
continuum.	

\paragraph{Aromatic band ratios in nearby galaxies}
\citetalias{galliano08b} studied the 6.2, 7.7, 8.6 and $11.3\mmic$ \hUIB s,
in a sample of nearby galaxies and Galactic regions.
Several band ratios were used in order to solve the degeneracies between
the effects of the charge, size distribution, \hISRF\ hardness and extinction.
The \hLint{6.2}/\hLint{11.3} and \hLint{7.7}/\hLint{11.3} ratios vary by a 
factor of $\simeq10$.
Among this sample, covering a large range of physical conditions and spatial 
resolutions, it appears that most of the variations of the band ratios arise 
from variations of the \hPAH\ charge.
\citetalias{smith07} also found large variations of the band ratios (by a 
factor of $\simeq2-5$) in central regions of starbursts and \hAGN s.
However, they found that the presence of a low-luminosity \hAGN\ could alter 
the spectrum, by destroying the smallest \hPAH s
\citep[also confirmed by][]{sales10}.
Harsh environments result in selective destruction of the smallest \hPAH s, 
like in elliptical galaxies \citep{kaneda07,vega10}, or in the superwind 
of \M{82} \citep{beirao15}.

In low-metallicity systems, the variations can be more difficult to probe, as
the band equivalent widths are lower (\refpanel{fig:MIR}{a}; this point is 
discussed in \refsecs{sec:destPAH}{sec:formPAH}).
In the \hLMC, \cite{mori12} found different trends in neutral and ionized 
sightlines.
Toward the latter, there are evidences that \hPAH s have a lower charge, as a 
consequence of the higher recombination rate, and are on average larger, 
due to the destruction of the smallest ones.
In contrast, in the \hSMC, \citet{sandstrom12} found very weak 
\hLint{6-9}/\hLint{11.3} ratios and weak 8.6 and $17\mmic$ bands, implying 
small weakly ionized \hPAH s.
This last point is consistent with the trend of \hLint{17}/\hLint{11.3} with 
\hmet\ found by \citetalias{smith07}.
However, \citet{hunt10} argued that Blue Compact Dwarf galaxies (\hBCD) exhibit a deficit of small \hPAH s.
If there is a smooth variation of \hPAH\ size distribution with \hmet, these 
results are in contradiction.
\cite{sandstrom12} noted that these \hBCD s are more extreme environments than
the \hSMC, and that photodestruction could dominate the \hPAH\ processing
(\cf~\refsec{sec:destPAH}).
We note that the solution to this apparent controversy might alternatively 
reside in the difference in studied spatial scales.
In the Magellanic Clouds, \hspitz\ spectroscopy gives a spatial resolution of a 
few parsecs, compared to a few hundreds in nearby \hBCD s.
The fact is that the \hLMC\ and \hSMC\ exhibit strong spatial variations of 
their \hUIB\ spectrum.
\citet{whelan13} showed a diversity of \hMIR\ spectral properties in the \hSMC.
They demonstrated that the \hPAH\ emission in a region like N$\,$66 is 
dominated by its diffuse component, and not by its bright clumps, where \hPAH s 
are destroyed.
At the other extreme, the molecular cloud SMC-B1\#1 shows unusually high \hUIB\ 
equivalent widths \citep{reach00}.
Also, the \hLint{11.3}/\hLint{12.7} ratio indicates that \hPAH s are more 
compact in \xxxdor\ and more irregular outside \citep{vermeij02}.
All these elements suggest that there is a complex balance of processes shaping 
the \hMIR\ spectra throughout low-\hmet\ environments.

Finally, \hUIB\ ratios can be used as a diagnostic tool of the physical 
conditions.
For instance, \citetalias{galliano08b} provided an empirical calibration of the 
\hLint{6.2}/\hLint{11.3} ratio with the \hUV-field-to-electron-density ratio, 
$G_0/n_e$ (\refsec{sec:PDR}).
However, the dynamics of the \hUIB\ ratios being at most a factor of 
$\simeq10$, with typical uncertainties of $20\,\%$, their diagnostic potential
is practically limited to a low/high value dichotomy.

    \subsubsection{Silicate Features in Emission}
    \label{sec:silem}
    
Silicates are one of the most abundant dust species
\citepalias[$\simeq2/3$ of the mass;][]{draine07,jones17}.
They are characterized by two prominent features at 9.8 and $18\mmic$, 
corresponding to Si$-$O stretching and O$-$Si$-$O bending modes, respectively 
(\refpanel{fig:MIR}{a}).
These features are most often seen in extinction (\refsec{sec:MIRext}).
They can be seen in emission in galaxies, when the dust is hot enough 
($T\gtrsim150\,$K).
It is the case near the central engine of \hAGN s 
\citep{wu09,hony11}, 
even low-luminosity \hAGN s, like the nucleus 
of \M{31} \citep{hemachandra15}.
The \hMIR\ spectra of early-type galaxies also show clear silicate emission 
features, but likely of circumstellar origin \citep[\eg][]{bressan06}.
Alternatively, some dust models present a significant out-of-equilibrium 
emission from small silicates \citep[\eg][]{zubko04}.
Such small silicates are not unlikely (\eg\ \refsec{sec:AME}).
In the diffuse \hISM, the $9.8\mmic$ feature would be hidden by the bright 
aromatic features.
However, we should be able to see it when the \hUIB\ intensity decreases, like 
in dwarf galaxies.
It is usually not the case \citep[\eg][]{remy-ruyer15}.
It might suggest that either these small silicates have not abundantly formed, 
or that they are efficiently destroyed in dwarf galaxies.

  \subsubsection{Aliphatic Feature in Emission}
  
The $3.4\mmic$ aliphatic emission feature is carried by small \hHAC.
The \hLint{3.4}/\hLint{3.3} aliphatic-to-aromatic ratio shows regional 
variations in the \hISM, as the result of structural changes in the 
hydrocarbons through \hUV\ processing \citep[\eg][]{jones13}.
\citet{yamagishi12} detected this feature in the superwind of \M{82}.
They found that the \hLint{3.4}/\hLint{3.3} ratio increases with distance from 
the center.
They interpreted this trend as the production of small \hHAC, by shattering 
of larger grains in this harsh halo.
Similarly, \citet{kondo12} found a higher \hLint{3.4}/\hLint{3.3} ratio in the 
nuclear bar of \ngc{1097}, suggesting that the gas flow towards the center 
could lead to the formation of small \hHAC\ by shattering.
We note that, alternatively, the \hLint{3.4}/\hLint{3.3} ratio can increase 
with the accretion of \hHAC\ mantles in denser regions \citep{jones13}.
This feature can also be seen in extinction, in \hAGN s \citep[\eg][]{mason07}.

    \subsection{The Wavelength-Dependent Extinction}
    \label{sec:extinction}

  \subsubsection{General features}

\begin{figure}[h]
  \includegraphics[width=1.23\textwidth]{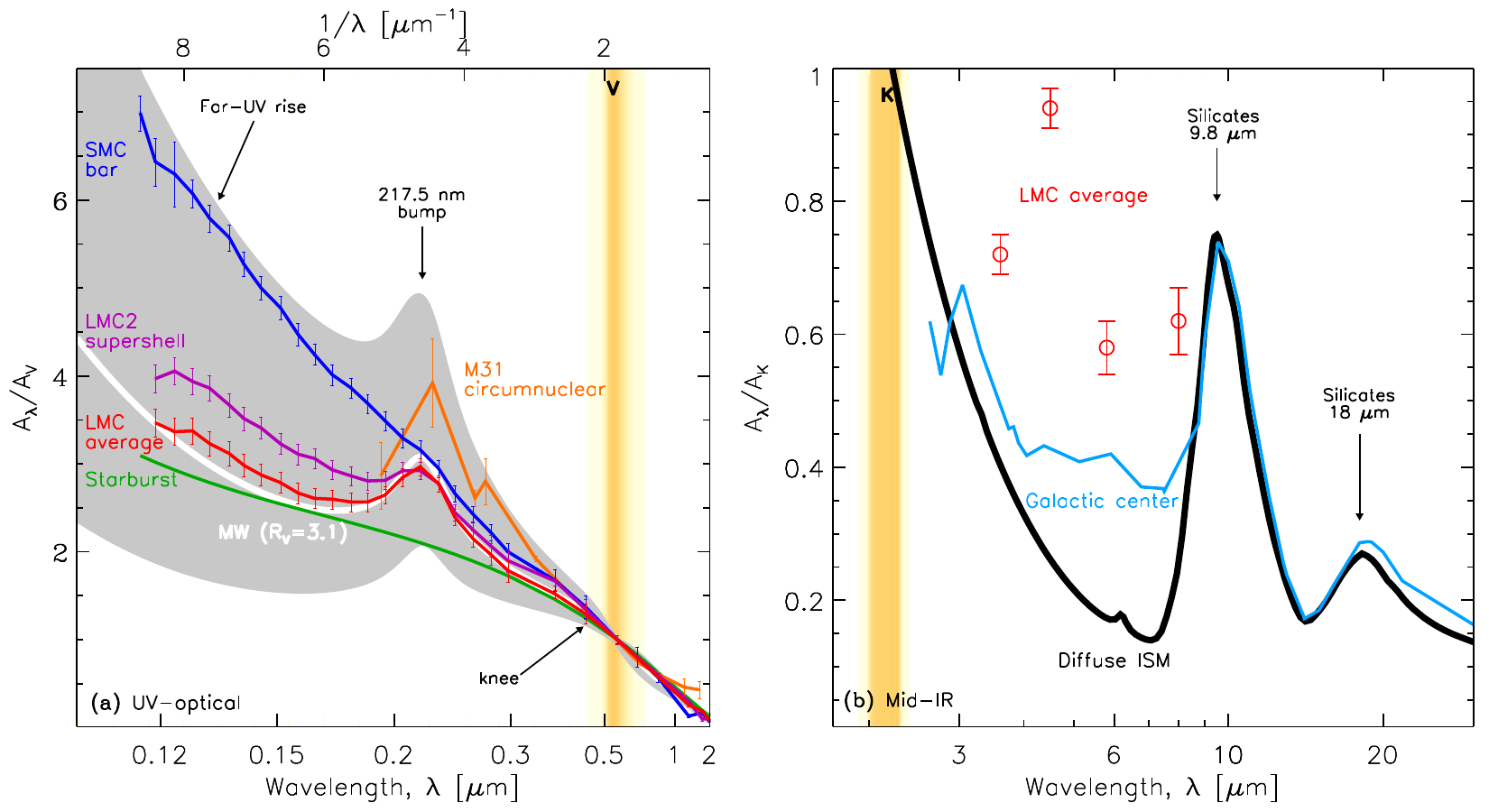}
  \caption{{\sl Extinction curves.} 
           \textitem{a} Extinction curves of the Magellanic clouds 
           \citep{gordon03}, compared to the \hMW 
           \citep[the grey area spans $\mRv=2-5$; the white curve corresponds 
           to $\mRv=3.1$][]{fitzpatrick99}.
           We have also displayed the attenuation curve of starburst galaxies
           \citep{calzetti00} and of the circ\-um\-nuclear region of \M{31} 
           \citep{dong14}.
           \textitem{b} \hMIR\ \hLMC\ average \citep{gao13},
           compared to the Galactic center \citep{lutz96}.}
  \label{fig:ext}
\end{figure}

The extinction is the combination of light absorption and scattering.
Early dust studies, before \hiras, were mainly based on extinction. 
In the \hMW, \citet[][\citetalias{cardelli89}]{cardelli89} demonstrated 
that the \hUV-to-\hNIR\ extinction curves follow a universal law, parameterized by 
\Rv:
\begin{equation}
  \mRv \equiv \frac{\mAv}{\mAb-\mAv}\;\;\;\mbox{ with }\;\;\;
  A_\lambda = 1.086
              \times\underbrace{\overbrace{\left(\kappa_\sms{abs}(\lambda)+\kappa_\sms{sca}(\lambda)\right)}^\sms{intrinsic grain opacity}
              \times\overbrace{\Sigma_\sms{dust}}^\sms{mass surface density}}_\sms{optical depth, $\tau(\lambda)$}.
\end{equation}
The sum of the absorption and scattering opacities, 
$\kappa_\sms{abs}+\kappa_\sms{sca}$, depends on the dust constitution.
Such an extinction curve presents different regimes (\refpanel{fig:ext}{a}):
\textitem{1}~a \hFUV\ rise, mainly due to absorption by small grains (Rayleigh 
limit: $A_\lambda\propto1/\lambda$);
\textitem{2}~a 217.5~nm absorption bump, probably carried by small carbon 
grains (\hPAH s, graphite, amorphous carbon, \etc);
\textitem{3}~an optical knee, mainly due to scattering by large grains;
\textitem{4}~a power-law \hNIR\ continuum.
On average, $\mRv\simeq3.1$ in the Galaxy, with large variations between 
sightlines ($\mRv\simeq2-5$) due to dust processing, low values of 
\Rv\ being attributed to regions with enhanced small grains.
For a given \Rv, the quantity $\mAv/\mN{H}$ is proportional to 
the dust-to-gas mass ratio.

    \subsubsection{Methodology}
    
The most reliable wavelength-dependent extinction curves are derived using the 
\citext{pair method} \citep{stecher65}.
Two stars of the same spectral type are observed, one with a low and one with a 
high foreground extinction.
The extinction curve is directly derived from the differential spectrum, 
assuming the dust properties are uniform along the two sightlines.
In external galaxies, this method has been successfully applied only to the
Magellanic clouds 
\citep[\eg][]{nandy81,gordon03} 
and \M{31} 
\citep{bianchi96b,clayton15}.
Alternatively, a stellar atmosphere model can be used in lieu of the reference 
star \citep[\eg][]{fitzpatrick05}.

In more distant objects, where observations of individual stars become 
impractical, other methods, less direct and more model dependent, have to be 
used.
Stellar \hSED\ modelling with dust attenuation is widely used 
\citep[\eg][]{hutton15}.
Attenuation, which is the net loss of photons within a galaxy, differs from 
extinction because it includes the effects of geometry.
There is a degeneracy between dust attenuation and 
stellar age and metallicity.
This degeneracy can be solved by accounting for \hFIR\ emission 
\citep{gordon00} or by studying disks with different inclinations 
\citep[][hereafter \citetalias{conroy10}]{conroy10}.
Alternatively, the ratio of Hydrogen recombination lines can be used.
\citet[][hereafter \citetalias{calzetti94}]{calzetti94} derived an average 
attenuation law 
(\refpanel{fig:ext}{a}) using this method, on a sample of 39 starbursts, 
assuming that all these objects had the same metallicity, stellar populations 
and dust properties.
This average curve is rather flat and has no bump (\refpanel{fig:ext}{a}).
In contrast, \citetalias{conroy10} derived an average attenuation law for 
disk galaxies, based on the UV-visible photometry.
They found that the \citetalias{calzetti94} law does not provide a good fit to 
their curve but that there might be a bump.
They point out that it is possible that previous studies failed to recognize 
the presence of the \hUV\ bump, because they tried \hMW\ laws only with 
$\mRv=3.1$.

Finally, other less common methods are available.
In a few rare cases, a background lenticular galaxy can be used to probe the 
extinction in a foreground galaxy \citep[\eg][]{white92,berlind97}.
Color magnitude diagrams can be used to probe the visible-\hNIR\ extinction 
curves, like the red giant clump \citep{gao13,de-marchi14}.
Supernova also provide bright sources probing the extinction in the host 
galaxy \citep[\eg][]{patat15}.

    \subsubsection{Extragalactic Results}

External galaxies help us probe the extinction curves in extreme conditions, 
beyond the simple $R_\sms{V}$ parameterization.
There is a continuous variation in shape of the extinction curves from the 
Galaxy to the \hSMC\ \citep[\refpanel{fig:ext}{a};][]{gordon03}.
The \hSMC\ bar is characterized by a very steep curve ($\mRv=2.74\pm0.13$), 
without 217.5~nm bump.
This is the sign of smaller grain size and reduced fraction of carbon, likely 
the result of grain shattering by supernova shock waves 
\citep{clayton03,cartledge05}.
A weak bump is seen in the \hSMC\ wing, which is more quiescent.
The \hLMC\ is intermediate between the Galaxy and the \hSMC, although there are 
important variations within.
The \hLMC$\,$2 supershell, near the massive star forming region \xxxdor, has 
a steeper law than the Galaxy ($\mRv=2.76\pm0.09$), with a weaker bump.
The average \hLMC\ is closer to the Galaxy ($\mRv=3.41\pm0.28$).
Neither the \hSMC, nor the \hLMC$\,$2 supershell conform to the 
\citetalias{cardelli89} parameterization.
There still remain some uncertainties on these properties.
\citet{de-marchi14} studied the visible-\hNIR\ curve in \xxxdor, using the 
spread of the red giant clump in the colour-magnitude diagram, and found a 
flatter extinction ($\mRv\simeq4.5$) that they 
attributed to freshly injected large grains.

The extinction curves of \M{31} ($Z=1-1.5\;Z_\odot$) are consistent with the 
\hMW\ \citep[][pair method]{clayton15}.
In its circumnuclear region, the curves appear steeper ($\mRv\simeq2.5$), but 
similar to the Galactic bulge \citep[][{$ $\hSED}]{dong14}.
In one of the regions studied by \citet{dong14}, there is a significantly
stronger \hUV\ bump (\refpanel{fig:ext}{a}).
In contrast, the spiral galaxy \M{101} appears to have a smaller bump, with a
similar continuum \citep[][{$ $\hSED}]{rosa94}.
In the edge-on starburst, \M{82}, \citet[][{$ $\hSED}]{hutton15} find that both \Rv\ 
and the bump strength decrease outward, with galactocentric distance.
The spiral galaxies studied with the overlap method all exhibit \hMW\ curves 
\citep[\eg][]{dewangan99,deshmukh13}, 
as well as early-type galaxies 
\citep[\eg][]{goudfrooij94,finkelman10}.
Some \hAGN s show signs of dust processing, with an absence of features 
\citep{crenshaw01,maiolino01}.

The extinction estimated towards \hSNII\ exhibit steep extinction curves
($\mRv\simeq1.5$), probably due to dust shattering by the shock wave 
\citep{hill95a,amanullah14,hutton15}.
The trend is the same towards \hSNIa, with more dispersed values 
\citep[$\mRv\simeq1.2-3$;][]{elias-rosa06,huang17}.

    \subsubsection{Silicate Features in Absorption}
    \label{sec:MIRext}

In the \hMIR, the extinction curve is dominated by the two prominent silicate 
features at 9.8 and 18~\mic\ (\refsec{sec:silem}; \refpanel{fig:ext}{b}).
In the Galaxy, $A_V/A_{9.8\mu{\rm m}}\simeq12$.
These absorption features can be seen when sufficient amount of material is 
obscuring a bright \hMIR\ source (like a star forming region or an \hAGN).
They are even seen at extremely low-metallicity \citep[like in SBS$\,$0335-052; 
$Z\simeq1/35\,Z_\odot$;][]{thuan99,houck04}.
One issue is the composition of these silicates.
It is well established that silicates are partially crystalline in 
circumstellar environments and become completely amorphous in the \hISM\ 
\citep[\eg][]{kemper04}.
The crystalline-to-amorphous ratio provides clues of dust processing, as 
crystallization and annealing have high energy barriers.
For instance, \citet{spoon06} detected distinctive crystalline silicate 
absorption features in several \hULIRG s.
Their crystalline-to-amorphous fraction is $\simeq10\,\%$.
They interpreted this value as the result of fresh injection of crystalline 
material by the massive star formation.
On average, observations of nearby galaxies find mostly amorphous silicates.
In a large sample of \hLIRG s and \hULIRG s, \citet{stierwalt14} reported the detection 
of crystalline silicate absorption features in only $6\,\%$ of their sample, in 
the most obscure sources.

    \subsubsection{Ice Absorption Features}
    
In shielded regions, some molecules can freeze out to 
form icy grain mantles.
The dominant species, H$_2$O, CO and CO$_2$ produce \hMIR\ absorption bands
(\refpanel{fig:MIR}{a}).
In the Magellanic clouds, ice absorption can be studied in individual young 
stellar objects \citep[\eg][]{oliveira13}.
In other galaxies, ice features likely come from molecular clouds.
\hULIRG s show particularly high ice optical depths, correlated with the 
silicates \citep{stierwalt14}.
However, some galaxies with high silicate absorption do not present detectable 
ice features, suggesting that the density of the medium is not the only 
parameter; the presence of an \hAGN\ could prevent ice formation.
In nearby star forming galaxies, \citet{yamagishi15} conducted an extensive
analysis of the CO$_2$-to-H$_2$O ice absorption ratio.
This ratio exhibits variations by a factor $\simeq20$, as H$_2$O has a longer 
lifetime (higher sublimation temperature) than CO$_2$.
They found this ice ratio correlates best with specific \hSFR\ (\hsSFR), indicating a high ratio 
in young star forming galaxies.

  \subsection{Elemental Depletion Patterns}
  \label{sec:depletion}

    \subsubsection{Definition \&\ Method}

Dust is made of the available heavy elements produced by stars.
In the \hISM, a fraction of these elements is in the gas and the rest is 
locked up into dust.
The fraction of missing elements from the gas is called \emph{depletion}.
In the \hMW, where depletions have been studied for most relevant heavy 
elements, in several hundreds of sightlines through the diffuse neutral medium
\citep{jenkins09}, it appears that:
\textitem{1}~\emph{refractory} elements, the elements with a higher 
condensation temperature ($T_\sms{c}$), are more depleted than \emph{volatile} 
elements, the elements with a lower $T_\sms{c}$ 
\citep[\eg][]{savage96};
\textitem{2}~the depletions increase the with mean density of the medium 
\citep[\eg][]{savage79b,crinklaw94}.
The depletion of an element X, is defined as:
\begin{equation}
  \underbrace{\left[\frac{\rm X_\sms{gas}}{\rm H}\right]}_\sms{depletion of X}
    \equiv 
  \underbrace{\log\left(\frac{\rm X}{\rm H}\right)_\sms{gas}}_\sms{abundance of X}
    - 
  \underbrace{\log\left(\frac{\rm X}{\rm H}\right)_\sms{ref}}_\sms{reference value}
    \simeq 
  \underbrace{\left[\frac{\rm X_\sms{gas}}{\rm H}\right]_0}_\sms{minimum depletion}
    + 
  \underbrace{A_\sms{X}\times F_\star}_\sms{local variation}.
  \label{eq:depletion}
\end{equation}
The second part of \refeq{eq:depletion} is a universal parameterization as a 
function of the depletion strength, $F_\star$ \citep{jenkins09}.
The minimum depletion term is thought to correspond to the core of the grain, 
while the varying environmental factor, $A_\sms{X}F_\star$, is attributed to 
accretion of mantles in denser environments.

Depletion measurements provide a direct estimate of the dust content, 
independent of model assumptions.
In addition, they provide constraints on the stoichiometry through the number 
ratio of available elements.
These measurements are however challenging.
In the Diffuse Neutral Medium (\hDNM), they are 
performed in absorption.
Most of the transitions are in the \hUV, thus inaccessible from the ground.
Apart from possible line saturation and ionization corrections, the most 
challenging aspect is the adoption of \emph{reference abundances} 
\refeqp{eq:depletion}, representing the total \hISM\ (gas and dust) abundances.
Indeed, the different standards (solar, meteoritic, F \&\ G stars, \etc) 
exhibit some inconsistencies.
Alternatively, \hii\ region abundances can be estimated from nebular lines.
These estimates rely on photoionization modelling, which add up another layer 
of uncertainties.

    \subsubsection{Extragalactic Depletions}

In extragalactic systems, there are numerous studies of \hDNM\ depletions in 
damped Ly-$\alpha$ systems \citep[{\hDLA}; \eg][]{de-cia16} or in $\gamma$-ray 
bursts \citep[\eg][]{friis15}, facilitated by the redshifting of the 
transitions in the visible range. 
Although such studies are usually compelled to make corrections based on the 
\hMW, \citet{de-cia16} adopted a holistic approach and derived 
self-consistent depletion sequences.

The depletions of several nearby galaxies have been studied.
The most complete studies concern the Magellanic clouds, with several 
elements, along several individual sightlines 
\citep[\eg][]{roth97,welty97,sembach01,sofia06,tchernyshyov15,jenkins17}.
Reference abundances can be estimated \insitu, directly from individual stars 
\citep[\eg][]{welty97,tchernyshyov15}.
The main results are the following.
\textitem{1}~The \hLMC\ depletion pattern of most elements appears 
Galactic, the abundances being simply scaled down by the metallicity 
\citep{tchernyshyov15}.
\textitem{2} In the \hSMC, there is a larger dynamics, likely resulting from 
more important processing (removal by shock waves). 
In addition, Si, Cr and Fe are systematically less depleted, with 
$\rm[Si/H]_0$ consistent with 0 \citep{tchernyshyov15}.
However, \citet{jenkins17}'s $\rm[Si/H]_0$ is significantly larger than 0.
Si depletion varies by a factor of $\simeq2$ at nearly constant Fe depletions, 
suggesting a different condensation process \citep{tchernyshyov15}.
\textitem{3} The C/O ratio is a factor of $\simeq2$ below solar, in the 
Magellanic clouds, suggesting that the carbon-to-silicate grain mass  
fraction is lower (\reftab{tab:depletions}).
\textitem{4} The Magellanic stream has a similar depletion pattern to the 
\hSMC\ \citep{sembach01}.

\begin{table}[htbp]
  \caption{Abundances and depletions in the Milky Way 
           and the Magellanic clouds.}
  \label{tab:depletions}
  \begin{tabularx}{\textwidth}{l|*{5}{X}||r}
    \hline
    \hline
      \rowcolor{coltab1}
      X  & C & O & Mg & Si & Fe & Mass ratio \\
    \hline
      \rowcolor{coltab2}\multicolumn{7}{c}{Milky Way} \\
    \hline
      $10^{6}(\rm X/H)_\sms{ref}$ & $290_{-20}^{+30}$ & $580_{-60}^{+70}$ & $42_{-2}^{+2}$ & $41_{-2}^{+2}$ & $35_{-2}^{+2}$ & $Z_\odot=1/75$ \\
      $[\rm X_\sms{gas}/H]_0\;(\%)$ & $23_{-23}^{+27}$ & $2.3_{-2.3}^{+12.3}$ & $46_{-4}^{+3}$ & $40_{-5}^{+5}$ & $89_{-1}^{+1}$ & $Z_\sms{dust}^{(F_\star=0)}=1/330$ \\
      $[\rm X_\sms{gas}/H]_0+A_\sms{X}\;(\%)$ & $39_{-11}^{+9}$ & $42_{-8}^{+7}$ & $95_{-1}^{+1}$ & $96_{-1}^{+1}$ & $99_{-1}^{+1}$ & $Z_\sms{dust}^{(F_\star=1)}=1/140$ \\
    \hline
      \rowcolor{coltab2}\multicolumn{7}{c}{LMC} \\
    \hline
      $10^{6}(\rm X/H)_\sms{ref}$ & $87_{-19}^{+25}$ & $320_{-70}^{+90}$ & $18_{-3}^{+4}$ & $22_{-5}^{+6}$ & $21_{-4}^{+4}$ & $Z=1/2\;Z_\odot$ \\
      $[\rm X_\sms{gas}/H]_0\;(\%)$ & $\unc{\simeq31}$ & $\unc{\lesssim20}$ & \ldots & $52_{-3}^{+3}$ & $89_{-1}^{+1}$ & $Z_\sms{dust}^{(F_\star=0)}=1/750$ \\
      $[\rm X_\sms{gas}/H]_0+A_\sms{X}\;(\%)$ & \ldots & \ldots & \ldots & $94_{-1}^{+1}$ & $99_{-1}^{+1}$ & $Z_\sms{dust}^{(F_\star=1)}=1/250$ \\
    \hline
      \rowcolor{coltab2}\multicolumn{7}{c}{SMC} \\
    \hline
      $10^{6}(\rm X/H)_\sms{ref}$ & $33_{-7}^{+9}$ & $140_{-20}^{+30}$ & $7.6_{-1.1}^{+1.2}$ & $9.1_{-2.0}^{+2.7}$ & $7.8_{-1.4}^{+1.6}$ & $Z=1/5\;Z_\odot$ \\
      $[\rm X_\sms{gas}/H]_0\;(\%)$ & $\unc{\simeq28}$ & $\unc{\simeq32}$ & $49_{-6}^{+6}$ & $40_{-3}^{+3}$ & $89_{-1}^{+1}$ & $Z_\sms{dust}^{(F_\star=0)}=1/2760$ \\
      $[\rm X_\sms{gas}/H]_0+A_\sms{X}\;(\%)$ & \ldots & \ldots & $71_{-18}^{+11}$ & $95_{-1}^{+1}$ & $99_{-1}^{+1}$ & $Z_\sms{dust}^{(F_\star=1)}=1/630$ \\
    \hline
  \end{tabularx}
  \begin{tabnote}
    \textitem{MW}~solar abundance compilation and
    depletions from \citet{jenkins09}.
    \textitem{LMC}~stellar abundance compilation and depletions
    from \citet{tchernyshyov15};
    $\rm[C/H]$ and $\rm[O/H]$ from \citet{korn02}.
    \textitem{SMC}~stellar abundance compilation
    from \citet{tchernyshyov15}; 
    depletions from \citet{jenkins17};
    $\rm[C/H]$ from \citet{pena-guerrero12a}'s \hii\ modelling
    of \ngc{456};
    $\rm[O/H]$ derived from \citet{mallouris03}'s $\rm[O/Zn]$
  (Sk$\,$108, $\log N(\mbox{\hi})=20.5\Rightarrow F_\star\simeq0$)
    and \citet{jenkins17}'s $\rm[Zn/H]$.
   Values in grey indicate an estimate from nebular emission lines.
  \end{tabnote}
\end{table}  


Other more extreme objects, like the lowest metallicity nearby galaxy \izw\ 
\citep{aloisi03,lebouteiller13}, or several star forming galaxies 
\citep{james14} have been studied in absorption, with some uncertainties due to multiple sightlines with different physical conditions.
Several studies based on \hii\ emission lines focused on low-\hmet\
galaxies.
It appears that $\rm[Mg/O]$ and $\rm[Mg/S]$ correlate with \hmet\
\citep{guseva13}, implying a higher Mg depletion at high \hmet.
Similarly, $\rm[Fe/O]$ correlates with \hmet\ 
\citep[\eg][]{rodriguez-ardila05,izotov06}.
These results are qualitatively consistent with the Magellanic cloud trends
(\reftab{tab:depletions}).

    \subsection{Polarization Studies}
    \label{sec:pola}
    
Grains are one of the main agents polarizing light in galaxies.
Historically, \citet{hall49} and \citet{hiltner49} first noted that starlight 
was sometimes polarized to a few percents, and that the degree of polarization 
was correlated with the reddenning.
\citet{davis51} proposed that this polarization was caused by non-spherical 
grains.
\citet{stein66} predicted that the emission by such elongated grains should 
also produce polarized \hIR\ emission.
\citet{dolginov76} proposed radiative torques as the main mechanism of grain 
alignment.
Recently, the whole sky polarization map of the Galaxy, at 353~GHz ($1^\circ$ 
resolution), has been observed by \hplanck.
The \hsubmm\ polarization fraction can be as high as $20\,\%$ in the most 
diffuse regions ($\mN{H}\simeq 10^{20}\;\rm H\,cm^{-2}$) and decreases along 
the densest sightlines  
\citep[$\mN{H}\simeq 10^{22}\;\rm H\,cm^{-2}$;][]{planck-collaboration15}.
In the low density regime, this behaviour is consistent with turbulently 
disordered magnetic field orientations.
In regions of higher densities, more opaque, the trend can be explained by the 
decrease of the efficiency of radiative torque alignment.

In nearby galaxies, polarization studies have been conducted at all wavelengths 
from the \hFUV\ to the radio.
Dust-induced polarization is usually used to trace one of the three following 
phenomena.

\begin{textbox}[h]
  \sectbox{DUST-INDUCED POLARIZATION PROCESSES}%
  \subsubsectbox{Scattering}%
    The light from a bright source (star or \hAGN) scattered onto grains is
    partially polarized.
    The resulting fraction of linear polarization is a complex function
    of the dust composition, size and spatial distribution.
    This process is usually efficient in the \hUV\ 
    \citep[\eg][]{zubko00}.
    It applies to spherical and elongated grains, indifferently.
    Most radiative transfer models include this polarization mechanism.
  \subsubsectbox{Dichroic extinction}%
    The light from a background source seen through a cloud of 
    magnetically-aligned elongated grains is partially polarized,
    parallel to the magnetic field.
    In the \hMW, the wavelength-dependent linear polarization fraction,
    caused by this process, follows the \citet{serkowski75} law.
    It is efficient from the Near-\hUV\ (\hNUV) to the \hNIR, peaking around 
    $\lambda\simeq0.55$~\mic.
  \subsubsectbox{Dichroic emission}%
    The thermal emission from magnetically-aligned elongated grains is 
    polarized, perpendicular to the magnetic field.
    The wavelength-dependent fraction of linear polarization is rather flat
    over the whole \hFIR-to-\hsubmm\ regime \citep[\eg][]{guillet17}.
\end{textbox}

      \subsubsection{The Geometry of Complex Regions} 
      
Polarization can provide information on the geometry of unresolved or poorly 
constrained sources.
Most noticeably, the polarimetric observations of the central region 
of the archetypal Seyfert~2 galaxy, \ngc{1068}, were
interpreted as the scattering of light from the central accretion disk, 
obscured by a dusty torus \citep{antonucci85}.
These results were a major step towards the unified \hAGN\ model,
attributing the differences in observed properties of Seyfert galaxies to the 
difference in orientation of the central source \citep{antonucci93}.
Apart from \hAGN, the \hNIR\ polarimetry of \M{82} has contributed revealing 
a central nuclear star forming ring \citep{dietz89}.

Polarization studies have also shed light on the nature of 
diffuse \hUV\ halos of star forming galaxies: are those made of purely 
scattered light, or is there a faint stellar population?
The observations of edge-on galaxies, like 
\ngc{3125} \citep{alton94} or \M{82} \citep{yoshida11} have revealed that 
their dusty outflows were scattering light from the central starburst.
\citet{cole99} also studied the polarization of star forming regions in the 
\hLMC\ to constrain the extent of their \hUV\ halos.

      \subsubsection{The Structure of the Magnetic Field}

In case of dichroic extinction or emission, the polarization angle provides
a map of the projected magnetic field.
Such studies were performed in the \hNIR/visible, among others, 
in \ngc{891} \citep{montgomery14}, \M{82} \citep{greaves00,jones00},
the \hSMC\ \citep{lobo-gomes15}, \ngc{6946} \citep{fendt98}
and the dust lane of \cena\ \citep{scarrott96}.

      \subsubsection{The Dust Composition}

The wavelength-dependent shape of the polarization fraction provides valuable 
information on the dust constitution.
In particular, studies of \M{31} \citep{clayton04},
\M{82} \citep{kawabata14}, the \hLMC\ \citep{clayton96} and
the \hSMC\ \citep{rodrigues97} all concluded that the polarizing grains were 
smaller in these objects than in the \hMW.
Polarimetry can also be used to discriminate models.
For instance, \citet{mason07} studied the polarization of the 3.4~\mic\ 
aliphatic band in order to test the silicate core/organic mantle model 
\citep{li02b}.
They observed the interstellar medium of \ngc{1068} in absorption towards 
its central engine.
If the carriers of the 3.4~\mic\ band were onto large grains, the feature
would be significantly polarized, which is not the case.

  \subsection{Dust-Related Epiphenomena}
  
    \subsubsection{Dust Observables in the X-Rays}
    \label{sec:X}

Dust absorbs and scatters X-rays.
First, the photoelectric absorption edges of elements locked-up in grains 
contain potential information on their chemical structure 
\citep[\eg][]{lee09b}.
For instance, \citet{zeegers17} studied the Si K-edge along the line of sight 
of a Galactic X-ray binary.
They were able to constrain the column density and the chemical composition of 
the silicate grains.
Second, X-ray scattering halos can be used to constrain dust models 
\citep[\eg][]{smith16}.
As an illustration, \citet{corrales15} modelled the X-ray halo around 
Cygnus~X-3.
They were able to put constraints on the size distribution, especially on the 
large grain cut-off.
Lastly, \citet{draine04b} made a case that time-varying X-ray halo could be 
used to estimate distances of nearby galaxies, down to $1\,\%$ accuracy
for \M{31}.

Among external galaxies, the dust-scattering X-ray halos of several 
$\gamma$-ray bursts have been observed 
\citep[\eg][]{evans14}.
However, to our knowledge, no such study has been conducted in nearby galaxies.

    \subsubsection{The Diffuse Interstellar Bands}
    \label{sec:DIBs}

\hDIB s are ubiquitous absorption features in the $\simeq0.4-2$~\mic\ range.
Over 400 of them have been detected in the \hISM\ \citep{hobbs09}, since their 
discovery, a century ago \citep{heger22}.
They remain largely unidentified, although four of them have been attributed to 
C$_{60}^+$ 
\citep{campbell15}.
They are associated to dust, as their strength correlates with $E({B-V})$ 
at low values, but they disappear in denser sightlines \citep[\eg][]{lan15}.
To first order, \hDIB s correlate with each other, but there are some notable 
differences, suggesting that they have different carriers \citep{herbig95}.
For instance, the so-called C$_2$ \hDIB s \citep{thorburn03} appear to be found 
preferentially in diffuse molecular clouds.

\hDIB s are abundantly observed in external galaxies, including distant 
objects, like \hDLA s 
\citep[\eg][]{lawton08}.
As a narrow spectral feature, they can easily be separated in velocity from the 
foreground Galactic features.
They are observed in the \hISM\ of nearby galaxies along the lines of sight of:
\textitem{1}~stars, in \M{33} \citep{cordiner08}, \M{31} 
\citep{cordiner11}
and the Magellanic clouds 
\citep[\eg][]{welty06,van-loon13,bailey15}, 
among others;
\textitem{2}~SN$\,$II, in \cena\ \citep{phillips87,dodorico89}, 
\M{66} \citep{bolte89}, the \hLMC\ \citep{vidal-madjar87}, 
\M{100} \citep{cox08},
\M{82} 
\citep[\eg][]{welty14}, 
among others;
\textitem{3}~SN$\,$Ia 
\citep[\eg][]{sollerman05,huang17}; 
and
\textitem{4}~lenticular galaxies \citep{ritchey15}.

In solar metallicity, normal galaxies, the strengths of the \hDIB s and their 
relation to \Av\ seem to be similar to those of the \hMW\ 
\citep{heckman00,sollerman05,cordiner11,huang17}, although some differences can 
be found 
\citep[\eg\ in \M{100}, \M{33} and \M{82};][]{cox08,cordiner08,welty14}.
In contrast, the largest deviations are found in the Magellanic clouds.
\citet{welty06} reported that the \hDIB s are weaker in the \hLMC\ and \hSMC, 
compared to the \hMW, by factors of $\simeq7-9$ and $\simeq20$, 
respectively.
However, they found that the C$_2$ \hDIB s have the same strength per H atom as 
in the Galaxy.
\hDIB\ spectra appear to be controlled by the \hUV\ field intensity 
\citep{cox06}, with disappearance of the features in the ionized gas or in high 
\hUV\ field 
\citep{van-loon13,bailey15}.
\hDIB\ strength also scales with metallicity, due to both lower shielding and 
lower elemental abundances \citep{cox07,bailey15}.
Finally, \hDIB s appear to be linked with the shape of the extinction curve.
\citet{cox07} demonstrated that the sightlines with weak or non-existent 
2175~\AA\ bump are those with weak or non-existent \hDIB s, in the \hSMC.

    \subsubsection{Grain Photoluminescence}
    \label{sec:ERE}

Photoluminescence is a non-thermal emission process in which, subsequently to 
the absorption of a \hUV\ photon, a grain is brought to an excited electronic 
state.
After partial internal relaxation, a redder photon is emitted, bringing the 
electron back to its fundamental state.
The \hERE, which is a broad emission band, found in the
$\simeq0.6-0.9$~\mic\ range of a diversity of Galactic environments, 
is attributed to dust photoluminescence
\citep[\eg][]{witt04b}.
The nature of its carriers is still debated.
  
In nearby galaxies, \hERE\ has been spectroscopically detected
in the star forming region \xxxdor\ \citep[{\hLMC};][]{darbon98}, 
in \M{82} \citep{perrin95} and \ngc{4826} \citep{pierini02}.
In the dwarf galaxies \sbs\ and \ngc{4449}, \citet{reines08} and 
\citet{reines08a} modelled the photometric observations of several Super Star Clusters (\hSSC).
Their result exhibit a significant \hI\ excess that they attributed to 
\hERE.  
However, \citet{reines10} admitted that this excess could be accounted for by 
continuum and line emission of the ionized gas, in \ngc{4449}.

    \subsubsection{The NIR Excess}
    \label{sec:NIRex}
      
An excess emission above the extrapolated stellar continuum is often detected 
in the \hNIR\ range \citep{joseph84,hunt92}.
It is seen in disk galaxies 
\citep[\eg][]{lu03,boquien11}
and dwarf galaxies 
\citep[\eg][]{vanzi00,smith09}.
It potentially hampers our ability to accurately estimate the stellar mass from 
\hNIR\ photometry.
This excess could be due to:
\textitem{1}~nebular emission: \bra\ line and free-free continuum 
  \citep[\eg][]{smith09};
\textitem{2}~hot equilibrium dust \citep[\eg][]{vanzi00}, probably in 
  circumstellar disks \citep{wood08};
\textitem{3}~out-of-equilibrium small grains 
  \citep[\eg][]{boquien10}.

Studying disk galaxies, \citet{lu03} describe this excess, having a color 
temperature of $\simeq10^3\,$K and an intensity of a few percent of the total 
\hFIR.
They found this excess to correlate with the \hUIB\ intensity.
In general, this excess correlates with \hSFR\ indicators
\citep{boquien10,mentuch10}.
The emissivity of the three proposed phenomena listed above correlates 
with star formation activity.
In a filament of the \hBCD, \ngc{1569}, \citet{onaka10} showed 
spectroscopically that the nebular emission can not account for the excess, 
thus favoring hot dust.

    \subsubsection{The Submm Excess}
    \label{sec:submmex}
    
An excess emission above the modelled dust continuum is often detected, 
longward $\simeq500\mmic$.
The most significant reports of this \citext{\hsubmm\ excess} can not be 
accounted for by (\cf~\reffig{fig:dustobs}): free-free, synchrotron and 
molecular line emission \citep[\eg][]{galliano03}.
The first occurence of such an excess was unveiled by \citet{reach95}, 
studying the \hcobe\ observations of the \hMW.
Their \hIR--\hsubmm\ \hSED\ could be fitted with a \hMBB\ 
($\hmbeta=2$; \cf~\refsec{sec:MBB}), and an additional $4-7$~K component.
A few years later, \citet{lisenfeld02} and \citet{galliano03} found a statistically 
significant excess in the dwarf galaxy \ngc{1569}, at $850\mmic$ and 1.3~mm.
Several subsequent studies confirmed the presence of an excess in other late-type 
galaxies 
\citep[\eg][]{dumke04,bendo06a,galametz09},
including the global \hSED s of the Magellanic clouds \citep{israel10,bot10}.
\hhersc\ and \hplanck\ opened the way to more detailed tests.
In particular, \citet{paradis12} showed that the 500~\mic\ excess becomes 
significant in the peripheral regions of the \hMW\ ($>35^{\circ}$), as well as 
towards some \hii\ regions.
Its relative amplitude can be up to $\simeq20\,\%$.
Spatially resolved observations of the \hLMC\ have shown that the 
$500\mmic$ excess varies up to $\simeq40\,\%$ in certain regions and 
is anticorrelated with the dust surface density \citep{galliano11}.
When resolved in non-barred spirals, the \hsubmm\ excess is primarily detected in 
the disk outskirts, thus at low-surface density \citep[\eg][]{hunt15}.
\begin{marginnote}
  \entry{COBE}{COsmic Background Explorer ($\lambda=12-5000\mmic$; 
               $1989-1993$).}
\end{marginnote}

\paragraph{Reality of the phenomenon}
First, we emphasize that, by definition, this excess is model-dependent. 
Different dust opacities lead to different amplitudes of the excess.
For that reason, probing this excess with models which are not based on 
realistic optical properties is a non-sense.
Second, the shape of the \hSED\ is well characterized in this regime.
It has been observed at different wavelengths, with different instruments.
It is still present with the latest \hhersc\ calibration \citep{dale17}.
In addition, reports of a deficit are very rare.
Finally, \citet{planck-collaboration11} showed that, while the \hsubmm\ 
excess in the integrated \hSED\ of the \hLMC\ was consistent with Cosmic 
Microwave Background (CMB) 
fluctuations, the \hSMC\ excess was significantly above this level.

\paragraph{Possible explanations}
The origin of the excess is currently debated.
The following explanations have been proposed.
\begin{description}
  \item[Very cold dust] (VCD) can be used to fit the excess.
    However, it leads to massive amounts of grains.
    \citet{galliano03} showed that VCD would be realistic only if this 
    component was distributed in a few number of dense, parsec-size clumps.
    Using the spatially resolved observations of the $500\mmic$ excess in the 
    \hLMC, \citet{galliano11} concluded that this explanation is unrealistic.
  \item[Temperature dependent emissivity.]
    The \citet{meny07} model predicts an increase of $\kappa(\lambda_0)$
    and a decrease of \hbeta\ with the temperature of amorphous grains. 
    It reproduces the \hMW\ excess \citep{paradis12} and the \hLMC\ 
    \citep[][coupled with spinning grains; \cf~\refsec{sec:AME}]{bot10}.
    However, it can not account for the excess in the \hSMC\ \citep{bot10}.
  \item[Magnetic grains.]
    \citet{draine12} showed that the \hSMC\ excess could be attributed to 
    magnetic nanoparticles (Fe, Fe$_3$O$_4$, $\gamma$-Fe$_2$O$_3$). 
    Thermal fluctuations in the magnetization of these grains can produce 
    strong magnetic dipole emission, since ferromagnetic materials are known to 
    have large opacities at microwave frequencies. 
    This hypothesis seems to be consistent with the observed elemental 
    abundances of the \hSMC\ and could also be responsible for the excess 
    detected in other environments. 
\end{description}

    \subsubsection{Spinning Grains}
    \label{sec:AME}
    
The \hAME\ is a cm continuum excess that 
can not be accounted for by the extrapolation of dust models, free-free, 
synchrotron and molecular line emission (\reffig{fig:dustobs}).
It was first detected in the \hMW\ \citep{kogut96b}.
\citet{draine98} promptly proposed that it was arising from the dipole emission 
of fastly rotating ultrasmall grains.
The candidate carriers were thought to be \hPAH s.
The \hwmap\ and \hplanck\ data of the Galaxy were successfully fit with 
spinning dust models, including \hPAH s 
\citep[\eg][]{planck-collaboration11e}.
In the \hMW, the \hAME\ correlates with all tracers of dust emission 
\citep{hensley16}.
However, \citet{hensley16} showed that \hAME/\hTIR\ ratio does not correlate 
with the \hPAH\ abundance.
These authors thus proposed that the carriers of the \hAME\ could be 
nano-silicates, rather than \hPAH s.
\begin{marginnote}
  \entry{WMAP}{Wilkinson Microwave Anisotropy Probe
               ($\lambda\simeq3.2-13$~mm; $2001-2010$).}
\end{marginnote}

In nearby galaxies, the first unambiguous detection of an \hAME\ has been 
obtained in an outer region of \ngc{6946} \citep{murphy10,scaife10}.
Follow up observations showed evidence for \hAME\ in 8 regions of this galaxy 
\citep{hensley15}.
This study showed that the spectral shape of this \hAME\ is consistent with 
spinning dust, but with a stronger \hAME-to-\hPAH-surface-density ratio, 
hinting that other grains could be the carriers.
Overall, the \hAME\ fraction is highly variable, in nearby galaxies.
\citet{peel11} put upper limits on the \hAME\ in \M{82}, \ngc{253} and 
\ngc{4945}.
These upper limits suggest that \hAME/100~\mic\ is lower than in the \hMW, in these objects.
In \M{31}, \citet{planck-collaboration15c} report a $2.3\sigma$ 
\citext{measurement} of the \hAME, consistent with the Galactic properties.
Finally, \citet{bot10}, fitting the \hNIR-to-radio \hSED\ of the \hLMC\ and 
\hSMC, 
temptatively explained the \hsubmm/mm excess with the help of spinning dust, in 
combination with a modified \hsubmm\ dust emissivity (\cf\ 
\refsec{sec:submmex}).
They conclude that if spinning grains are responsible for this excess, their 
emission must peak at 139~GHz (\hLMC) and 160~GHz (\hSMC), implying large 
\hISRF\ intensities and densities.
\citet{draine12} argued that such fastly rotating grains would need a \hPDR\
phase with a total luminosity more than two orders of magnitude brighter than 
the \hSMC.

  \addtocontents{toc}{\protect\pagebreak[4]}
  \section{EVIDENCE OF DUST EVOLUTION}

\begin{textbox}[h]
  \sectbox{DUST EVOLUTION PROCESSES}%
  
  \subsubsectbox{Grain Formation}%
    The dust mass is built up by:
    \textitem{1}~grain condensation in the ejecta of core-collapse \hSN e and 
      \hAGB\ stars.
    \textitem{2}~grain (re-)formation in the \hISM, by
	  accretion of atoms and molecules (grain growth, and mantle and ice 
	  formation).

  \subsubsectbox{Grain Processing}%
    The grain constitution is altered in the \hISM\ by:
    \textitem{1}~shattering and fragmentation by grain-grain collisions 
      in low-velocity shocks (modification of the size distribution);
    \textitem{2}~structural modifications by high energy photons or cosmic ray 
      impacts;
    \textitem{3}~coagulation.
    
  \subsubsectbox{Grain Destruction}%
  	The elements constituting the grains are partially or fully removed by:
	\textitem{1}~erosion and evaporation by thermal
	  or kinetic sputtering (gas-grain collision in a hot gas or a shock);
	\textitem{2}~photodesorption of atoms and molecules;
	\textitem{3}~thermal evaporation;
	\textitem{4}~astration (incorporation into stars).  
\end{textbox}

    \subsection{The Empirical Effects of Star Formation Activity \&\ 
                Metallicity}
    \label{sec:SFvsZ}

Dust evolution is the modification of the constitution of a grain mixture 
under the effect of environmental processing.
Most dust evolution processes can be linked to star formation:
\textitem{1}~formation of molecular clouds and their subsequent evaporation;
\textitem{2}~stellar ejecta;
\textitem{3}~\hSN\ shock waves;
\textitem{4}~\hUV\ and high-energy radiation.
The characteristic timescale of these processes is relatively short
(of the order of the lifetime of massive stars; $\tau\lesssim10$~Myr) and their 
effect is usually localized around the star forming region.
For these reasons, the \hsSFR\ is an indicator of sustained dust processing.
However, the dust lifecycle is a hysteresis.
There is a longer term evolution, resulting from the progressive elemental 
enrichment of the \hISM, which becomes evident on timescales of $\simeq1$~Gyr.
This evolutionary process can be traced by the metallicity.

\begin{figure}[h]
  \includegraphics[width=1.23\textwidth]{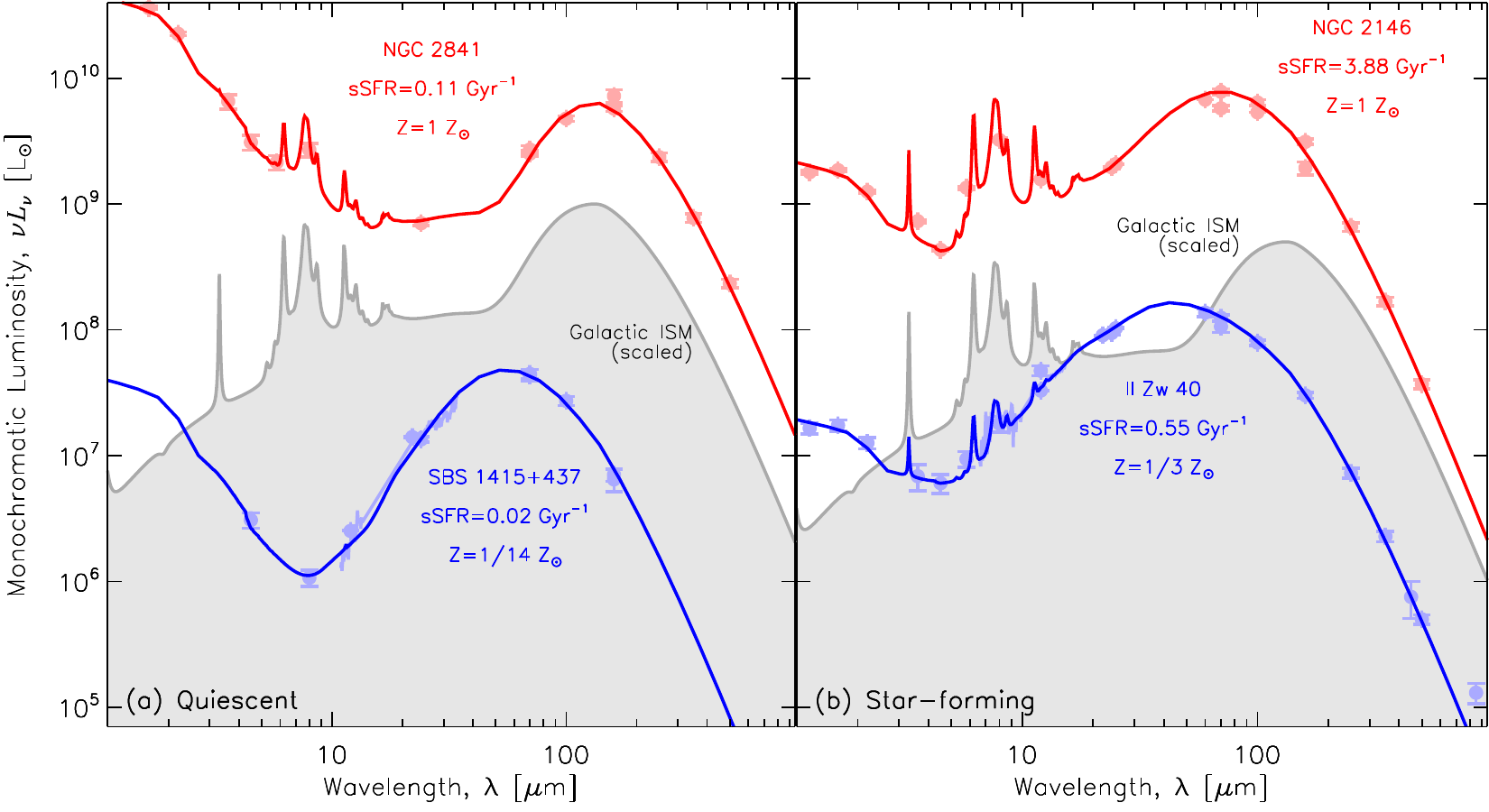}
  \caption{{\sl Effects of dust evolution on the \hSED s of galaxies.}
           Each panel displays the observations and the \hSED\ model of two
           nearby galaxies \citep{remy-ruyer15}, on top of the \hSED\ of the 
           diffuse Galactic \hISM\ (in grey).
           The red curve in panel \textitem{a} shows a quiescent solar
           metallicity galaxy. 
           Apart from the stellar continuum, it is identical to the diffuse
           \hISM.
           In contrast, the blue curve represents a low-metallicity quiescent
           system.
           Its dust properties are notably different: 
           \textitem{1}~weak or absent \hUIB s; 
           \textitem{2}~overall hotter dust (\hFIR\ peaks at shorter 
           wavelengths);
           \textitem{3}~a somehow broader \hFIR\ spectrum, resulting from a 
           distribution of starlight intensities and/or an overabundance of 
           small grains (\cf\ \refsec{sec:Udist}).
           This \hSED\ is qualitatively similar to the \hSED\ of a compact 
           \hii\ region \citep[\eg][]{peeters02}.
           The red curve in panel~\textitem{b} shows a solar metallicity
           galaxy with a sustained star formation activity.
           Compared to its quiescent counterpart, it has a much hotter and 
           broader \hFIR\ emission, originating at least partly in bright
           \hPDR s.
           The starbursting low-\hmet\ galaxy (blue curve in 
           \refpanel{fig:selectSED}{b}) has the same features as its quiescent 
           counterpart, with a broader \hFIR\ emission.}
  \label{fig:selectSED}
\end{figure}
These two evolutionary timescales have an impact on the integrated \hSED s of 
nearby galaxies.
\reffig{fig:selectSED} illustrates that the effects of star formation activity 
and metallicity are not always easy to disentangle.
Indeed, low-\hmet\ systems are often dominated by young stellar 
populations.
In addition, their lower dust-to-gas mass ratio renders their \hISM\ more 
transparent, and thus allows massive star formation to impact the \hISM\ in a 
larger volume (\cf~\refsec{sec:dwarf}).
In other words, some properties of dwarf galaxies may be the result of their
hard and intense radiation field, rather than their low metallicity.
In practice, dust evolution processes can be probed by comparing regions 
in a galaxy or by comparing integrated galaxies.
Our ability to observe \hISM\ dust evolution in real time is limited to
\hSN\ remnants \citep[\eg\ \hSN$\,$1987A;][]{dwek10}.

    \subsection{Localized Dust Processing}
    \label{sec:localevol}

      \subsubsection{Grain Growth}
      \label{sec:FIRevol}

There is clear evidence of \hFIR\ opacity variation in the \hMW.
The main factor seems to be the density of the medium.
For instance, \citet{stepnik03} found that the \hFIR\
dust cross-section per H atom increases by a factor of $\simeq3$ from the 
diffuse \hISM\ to the molecular cloud they targeted.
They noticed that this opacity variation is accompanied by the 
disappearance of the small grain emission.
They concluded that grain coagulation could explain these variations
\citep[see also][hereafter \citetalias{kohler15}]{kohler15}.
In the diffuse \hISM, \citet{ysard15} showed that the variation of emissivity, 
including the $\beta-T$ relation (\cf~\refsec{sec:MBB}), could be explained by 
slight variations of the mantle thickness of the \citetalias{jones17} model.
This observed behaviour is also consistent with the progressive de-mantling and 
disaggregation of molecular cloud-formed, mantled and coagulated grains 
injected into the low density \hISM, following cloud disruption. 
It is perhaps not unreasonable to hypothesise that dust growth in the \hISM\ 
occurs on short timescales during cloud collapse rather than by dust 
growth in the quiescent diffuse \hISM. 
In this alternative interpretation, the arrow of time is in the opposite sense 
and requires rapid dust growth, through accretion and coagulation, in 
dense molecular regions and slow de-mantling and disaggregation in the diffuse 
\hISM. 

In nearby galaxies, such studies are difficult to conduct, as the emissivity 
variations are smoothed out by the mixing of dense and diffuse regions.
Even, when potential evolutionary trends are observed, their interpretation is 
often degenerate with other factors.
The Magellanic clouds are the most obvious systems where this type of study
can be attempted.
The insights provided by depletion studies (\cf~\refsec{sec:depletion})
show that there are clear variations of the fraction of heavy elements 
locked-up in dust, and these variations correlate with the density 
\citep{tchernyshyov15,jenkins17}.
Since the coagulation and the accretion of mantles lead to an increase of
\hFIR\ emissivity \citepalias[\eg][]{kohler15}, we should expect emissivity 
variations in the Magellanic clouds.
Indeed, \citet{roman-duval17} studied the trends of gas surface density
(derived from \hi\ and CO) as a function of dust surface density (derived 
from the \hIR\ emission), in these galaxies.
They found that the observed dust-to-gas mass ratio of the \hLMC\ increases 
smoothly by a factor of $\simeq 3$ from the diffuse to the dense regions.
In the \hSMC, the same variation occurs, with a factor of $\simeq7$.
They argue that optically thick \hi\ and CO-free \hmol\ gas 
(\cf~\refsec{sec:gas}) can not explain these trends, and that
grain growth is thus the most likely explanation.
However, we note that the possible increase of the dust opacity with density 
could explain part of this trend.

      \subsubsection{Size Distribution Variations}

As we have seen in \refsec{sec:extinction}, there is a great diversity of
extinction curves in the \hMW\ and nearby galaxies.
A large part of these variations is thought to be the result of variations in 
the size distribution, small \Rv\ corresponding to an overabundance of small 
grains \citep[\eg][]{cartledge05}.
Comparing the \hSMC~bar to the \hSMC~wing or the \hLMC2 supershell to the 
\hLMC\ average (\reffig{fig:ext}), it appears that there are more small 
grains in regions of massive star formation.
In the same way, comparing the extinction curves in the \hMW, \hLMC\ and \hSMC, 
we also notice that there is a potential increase of the small grain fraction 
when the metallicity decreases.
As pointed out in \refsec{sec:SFvsZ}, the two effects are degenerate.

Constraining the size distribution from the \hIR\ emission is more difficult
due to the degeneracy between size and \hISRF\ distributions 
(\cf~\refsec{sec:Udist}).
However, \citet{lisenfeld02} and \citet{galliano03} attempted the modelling
of the \hIR\ \hSED\ of the dwarf galaxy \ngc{1569}, varying the size 
distribution.
They both concluded that the dust in this object is dominated by nano-grains.
Interestingly enough, the extinction curve corresponding to the grain 
properties of \citet{galliano03} was qualitatively similar to the \hSMC, with a 
steep \hFUV-rise and a weak bump.
\citet{galliano05} found the same result for three other dwarf galaxies.
In the \hLMC, \citet{paradis09} also concluded to a drastic increase 
of the fraction of very small grains, especially around \xxxdor.
It is possible that, even if a fraction of hot equilibrium dust has been 
mistaken for small grains by these studies, these systems harbor, on average, 
smaller grains than normal galaxies.
\hSN-triggered shock waves, which are abundant in star forming dwarf galaxies
(\cf~\refsec{sec:dwarf}), by fragmenting large grains, could 
explain the peculiar \hSED s and extinction curves of dwarf galaxies
\citep[\eg\ Fig.~17 of][$V_\sms{shock}=100\;{\rm km\,s^{-1}}$]{bocchio14}.

    \subsubsection{Grain Destruction}
    \label{sec:dest}

The dust destruction efficiency in \hSN-triggered shock waves was recently 
re-estimated using the \citetalias{jones17} model to evaluate 
the role of dust mantles, and to calculate the emission and 
extinction from shocked dust \citep{bocchio14}. 
Further constraints were put on the silicate destruction time, using
hydrodynamical simulations \citep{slavin15}.
The main conclusions of these studies are the following.
\textitem{1}~\hHAC\ grains are quickly destroyed, even in a 
$50\;\rm km\,s^{-1}$ shock, which is counter to earlier work 
\citep[\eg][]{jones96} that used the properties of graphite and an amorphous 
carbon other than \hHAC.
It implies that the re-formation of carbonaceous dust in the dense regions of 
the \hISM\ is a strong requirement (\cf~\refsec{sec:FIRevol}).
\textitem{2}~Silicate grains appear to be more resilient, with a mean lifetime
of $\tau_\sms{sil}\simeq2-3$~Gyr \citep{slavin15}, one order of magnitude
larger than the previous estimate of $\tau_\sms{sil}\simeq400$~Myr 
\citep{jones96}.

\paragraph{Photodestruction of small grains}
\label{sec:destPAH}
In Galactic \hPDR s, \citet[][Fig.~3]{boulanger98a} showed that
the \hUIB\ strength departs from a linear dependence on the \hISRF\ intensity,
\hU, for $\hmU\gtrsim1000$, indicating that band carrier destruction has 
occurred. 
It therefore implies that, in a galaxy with regions of intense 
stellar radiation, the observed \hUIB s are most likely not coming 
from high \hISRF\ regions. 
More recently, work on radiative transfer modelling of the dust emission in 
Galactic \hPDR s has shown that the small grains are significantly 
under abundant, with respect to the diffuse \hISM\ 
\citep{arab12}. 

This phenomenon is widely observed in nearby galaxies.
First, the \hUIB\ carriers appear cleared out of the hotter parts of 
extragalactic, star forming regions \citep[\eg][]{galametz13,wu15}.
This destruction impacts a larger volume in blue compact galaxies. 
For example, in \ngc{5253}, the $11.3\mmic$ equivalent width increases with 
the distance from the \hSSC\ region, over several hundreds of parsecs 
\citep{beirao06}.
Second, the $L_\sms{3.3}/L_\sms{IR}$ ratio of \hLIRG s and \hULIRG s decreases 
when $L_\sms{IR}$ increases 
\citep[\eg][]{imanishi10}.
Since $L_\sms{IR}$ quantifies the \hSFR\ (\cf~\refsec{sec:SFR}), this relation 
indicates that the \hUIB\ destruction can be seen at the scale of the whole 
galaxy.
Finally, there are well-known correlations between the \hUIB\ equivalent width
and various tracers of the intensity and hardness of the radiation field,
such as the ionic line ratio \neiiiline/\neiiline\ 
\citep{madden06,gordon08,lebouteiller11}.
These correlations are observed both within spatially resolved sources and 
among integrated galaxies.
These results suggests that scaling a diffuse \hISM-type dust emission to 
regions of enhanced radiation field, such as in \refpanel{fig:fitSED}{c}, is 
not appropriate.

\paragraph{Sputtering: grain evolution in hot plasmas}
The superwind of \M{82} exhibits filaments of dust and gas around the central 
outflows. 
\hPAH s or carbonaceous nano-particles embedded in such 
energetic regions would be exposed to soft X-rays ($0.5-2.0$~keV) and a hot gas 
($T_\sms{gas}\simeq10^7$~K; \refsec{sec:wind}).
In such regions, their survival time is only $\simeq20$~Myr, and that 
their destruction is principally due to collisions with the hot gas rather than 
by X-ray photo-destruction \citep{micelotta10b}. 
Yet, \hUIB s are detected in these outflows 
\citep[\eg][]{yamagishi12,beirao15}.
Thus, they are likely protected in the 
entrained cold gas that is being ablated into the hot outflowing gas, rather 
than present in the hot gas itself 
\citep{micelotta10b,bocchio12,bocchio13}.

    \subsection{Cosmic Dust Evolution}
        
      \subsubsection{Dust-Related Scaling Relations}
      \label{sec:scalrel}

\begin{figure}[h]
  \includegraphics[width=1.23\textwidth]{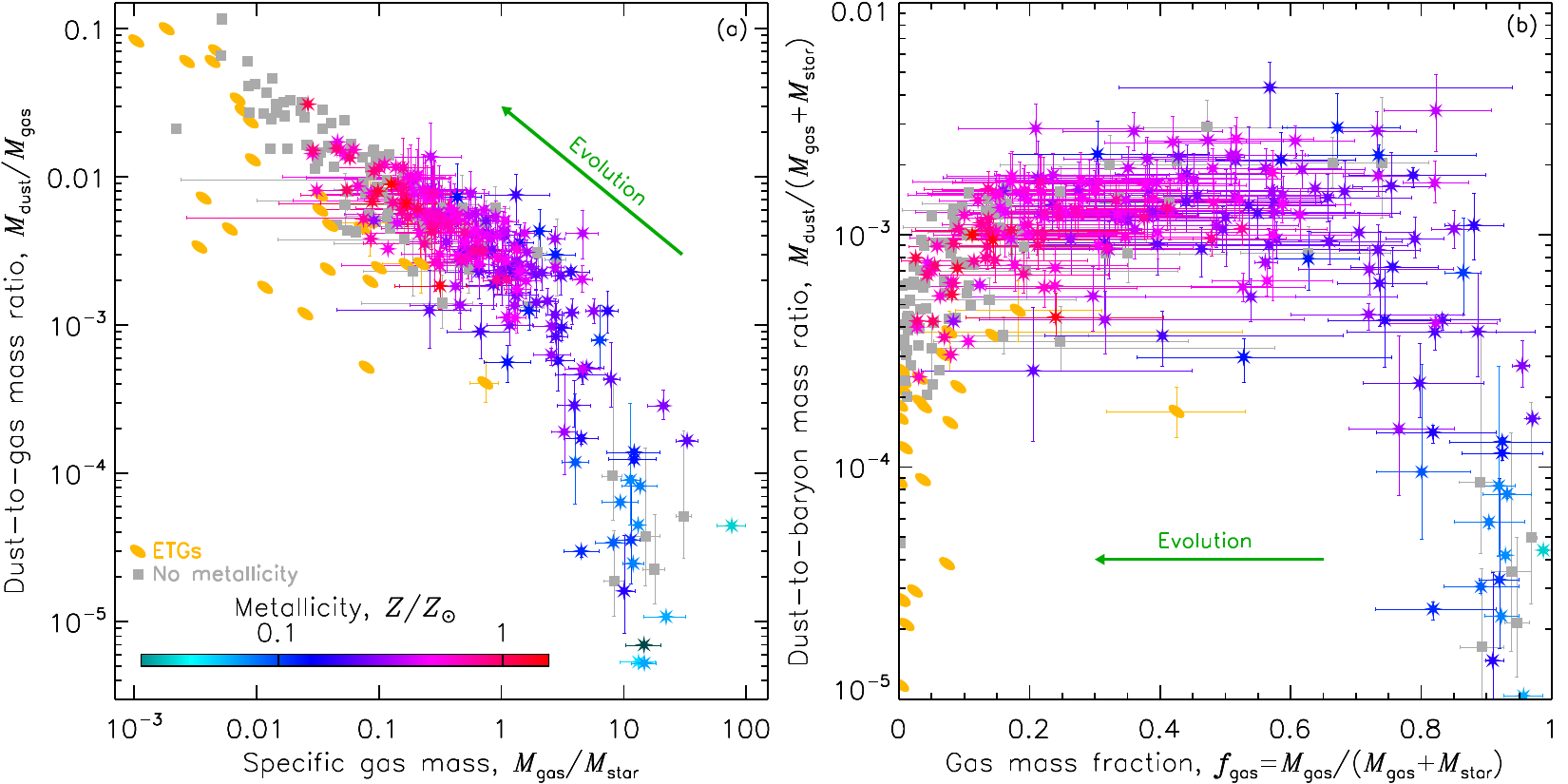}
  \caption{{\sl Main scaling relations.}
           Each point represents a nearby galaxy from the volume-limited
           samples of \citet[\hi-selected]{de-vis17a}, 
           \citet[$250\mmic$-selected]{clark15}, 
           \citet[$K$-band-selected]{cortese12} 
           and \citet[{\hmet}-selected]{remy-ruyer15},
           homogenized by \citet{de-vis17}.
           In both panels, the galaxies are color coded with metallicity, 
           when available.
           The gas mass is estimated with the sole neutral 
           atomic tracer 
           ($M_\sms{gas}\simeq 1.32M_\sms{\hi}$, accounting for He).
           }
  \label{fig:scalrel}
\end{figure}
The correlation between combinations of global parameters, such as 
$M_\sms{dust}$, $M_\sms{gas}$, $M_\sms{star}$ or \hSFR, across galaxy types, 
provides observational clues of cosmic evolution.
Dust-related scaling relations have thrived on \hhersc\ data, as this 
observatory has provided reliable dust mass estimates for statistical samples 
of galaxies, with various selection criteria 
\citep[\eg][]{cortese12,remy-ruyer15,clark15,de-vis17a}.
\refpanel{fig:scalrel}{a} displays the evolution of the dust-to-gas mass 
ratio, as a function of the specific gas mass.
When a galaxy evolves, its gas content is converted into stars, reducing
the $M_\sms{gas}/M_\sms{star}$ ratio.
At the same time, the \hISM\ is enriched in dust, increasing the 
$M_\sms{dust}/M_\sms{gas}$ ratio.
We also see that Early-Type Galaxies (\hETG), which are X-ray bright sources, 
lie notably below the correlation.
This is possibly the result of thermal sputtering of the grains in a hot 
plasma \citep[$T\gtrsim10^6$~K;][]{de-vis17a}.
\refpanel{fig:scalrel}{b} demonstrates the effect of astration on the dust 
content.
It shows the dust build-up, early-on ($f_\sms{gas}\gtrsim0.8$), then a 
plateau, and a net mass loss at ($f_\sms{gas}\lesssim0.2$).

      \subsubsection{Dust-to-Gas Mass Ratio Evolution with Metallicity}
      \label{sec:DG}

\begin{figure}[h]
  \includegraphics[width=1.23\textwidth]{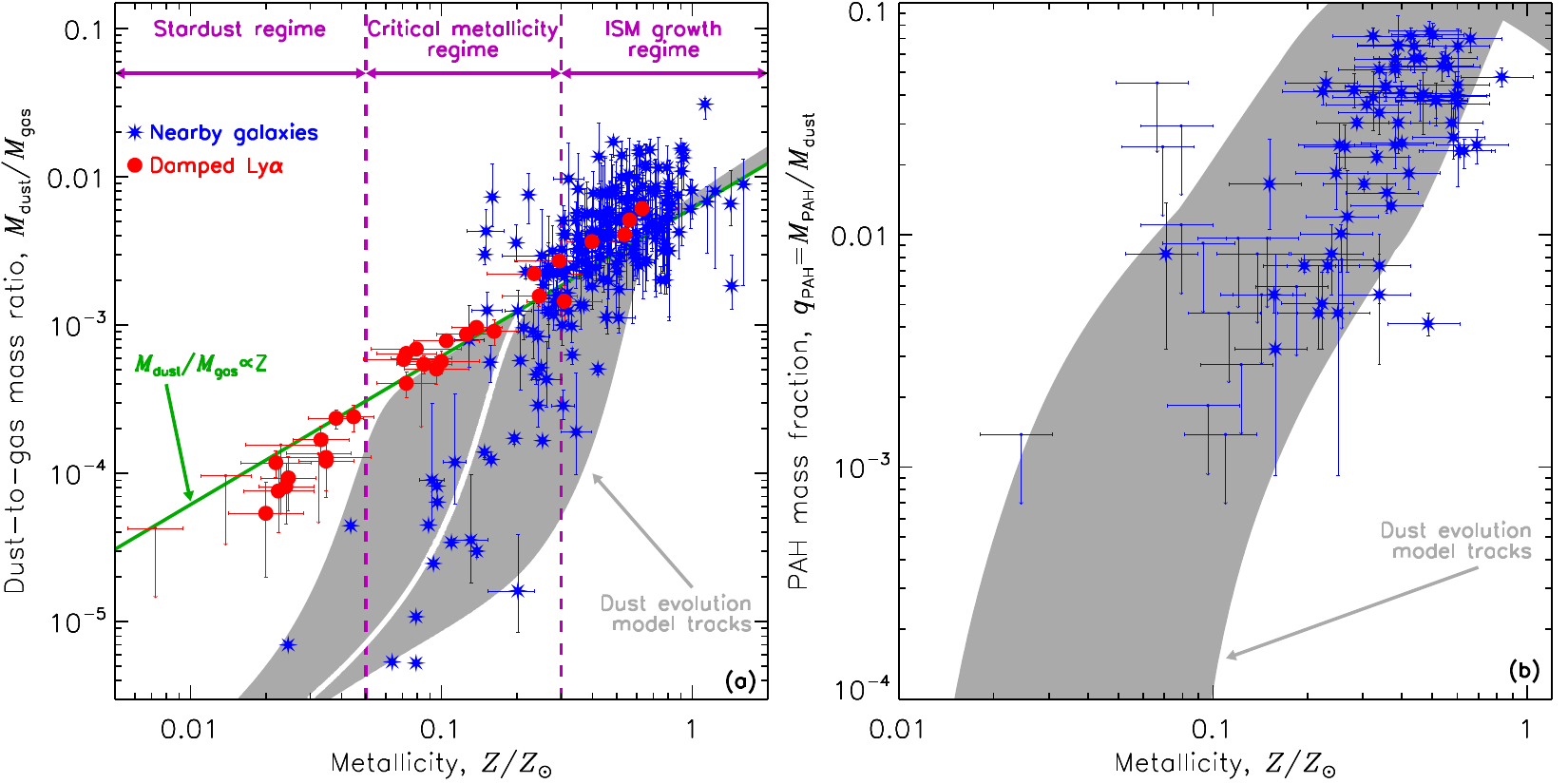}
  \caption{{\sl Dust evolution with metallicity.}
           \textitem{a}~Total dust-to-gas mass ratios of nearby galaxies
           \citep{de-vis17} and \hDLA s \citep{de-cia16}.
           The grey area represents the tracks of the model of \citet{asano13},
           varying the star formation timescale from
           $\tau_\sms{SF}=0.5$ to 50~Gyr ($\tau_\sms{SF}=5$~Gyr in white).
           The green line shows the locus of a constant, Galactic dust-to-metal 
           mass ratio.
           \textitem{b}~PAH-to-dust mass ratios
           for the nearby galaxy sample of \citet{remy-ruyer15}.
           The grey area shows the tracks of the \citetalias{galliano08a}
           dust evolution model.
           The metallicity, \hmet, has been derived from the O abundance.}
  \label{fig:dustevol}
\end{figure}
An important scaling relation, often used to constrain dust evolution models, 
is the trend of dust-to-gas mass ratio with metallicity 
\citep[\eg][]{lisenfeld98,draine07b,galliano08a,remy-ruyer14,de-vis17}.
\refpanel{fig:dustevol}{a} shows this relation for nearby galaxies (in blue).
It is clearly non-linear, suggesting that dust production is less 
efficient at early stages.
There are several sources of uncertainties that could bias this relation.
\textitem{1}~To derive the dust mass, the dust constitution has been assumed 
homogeneous throughout the whole sample.
However, the expected variations of the grain mixture constitution 
(\cf~\refsec{sec:FIRevol}) could alter the dust mass by a factor $\simeq2-3$ 
\citepalias{kohler15}, which is only a minor effect on a trend spanning four orders 
of magnitude.
\textitem{2}~The estimate of the gas mass could be inaccurate.
In particular, the absence of molecular gas mass constraints is problematic.
However, we note that \citet{remy-ruyer14} using CO-derived \hmol\ masses and 
exploring the effects of different CO-to-\hmol\ conversion factors, found a
similar trend.
As noted by \citet{de-vis17a}, the displayed sample is not \hmol-dominated, and 
the uncertainty on the total gas mass is not expected to be larger than a 
factor of $\simeq2$.
\textitem{3}~Low-metallicity galaxies usually have large \hi\ halos surrounding
their dust emitting region (\cf~\refsec{sec:dwarf}).
A dust-to-gas mass ratio encompassing the whole halo would therefore 
be underestimated \cite[see the discussion in][]{draine07b,remy-ruyer14}.
Although, this effect has been partially corrected for by \citet{remy-ruyer14}, 
based on the available \hi\ maps of their sample.
In particular, the \hi\ mass of the lowest metallicity source in
\refpanel{fig:dustevol}{a}, \izw, has been corrected.
It is therefore difficult to understand how the non-linearity of the trend 
would result from sole measurement biases.

We compare the nearby galaxy trend to the \hDLA\ sample of 
\cite{de-cia16}, in \refpanel{fig:dustevol}{a}.
The dust-to-gas mass ratio and the metallicity of \hDLA s are all estimated 
from redshifted \hUV\ absorption lines (\refsec{sec:depletion}).
These sources show an almost perfectly linear trend down to 
$\simeq1/100\;Z_\odot$. 
Their dust-to-metal mass ratio varies by only a factor of $\simeq3$ over the 
whole metallicity range.

      \subsubsection{Models of Global Dust Evolution}
      \label{sec:modevol}
      
It is possible to model the dust evolution of a galaxy, over 
cosmic time, by accounting for the balance between the production and 
destruction mechanisms.
This approach was initiated by \citet{dwek80}, who included grain processing
in gas enrichment models.
The main physical ingredients are the following.
\textitem{1}~The star formation history of the galaxy is the driving mechanism.
It can be parameterized, with different time-scales, episodic bursts or 
can be regulated by the inflow and outflow rates.
\textitem{2}~At a given time, different stellar populations are born (\hIMF\ 
dependent), they destroy a fraction of the dust by astration.
\textitem{3}~At the end of their lifetime, stars inject newly formed heavy 
elements and dust in the \hISM.
The stellar yields of core-collapse \hSN e and \hAGB\ stars are however quite
uncertain \citep[see the discussion in][]{matsuura15}.
\textitem{4}~Grains grow in the \hISM, by accretion. 
However, the associated sticking coefficients are uncertain.
\textitem{5}~Finally, as massive stars die, dust is destroyed by their 
\hSN-triggered shock waves, whose efficiency is also uncertain 
(\cf~\refsec{sec:dest}).
Most studies develop one-zone models.
The evolution of 
the size distribution can be tracked 
\citep[\eg][]{hirashita15a}.

Despite the noted uncertainty in the efficiency of the individual processes,
these models provide consistent trends of dust-to-gas ratio with 
metallicity, which enlighten the observations discussed in \refsec{sec:DG}.
We display the area covered by different dust evolution tracks in 
\refpanel{fig:dustevol}{a}.
These tracks correspond to different star formation histories.
We can see that they describe remarkably the nearby galaxy distribution.
They can be characterized by the three following regimes.
\textitem{1}~At low-\hmet, the grains are mainly condensed in stellar ejecta, 
the dust-to-gas ratio is proportional to metallicity, but with a low 
dust-to-metal ratio.
\textitem{2}~At intermediate metallicities, grain growth in the \hISM\ starts
to become important, as there are more heavy elements to be accreted.
The overall dust production efficiency increases.
\textitem{3}~At metallicity close to solar, we reach a linear regime, dominated
by grain growth.
The nearby galaxy trend in \refpanel{fig:dustevol}{a} therefore suggests that
grain growth is a crucial process (see also \refsec{sec:dest}).
The scatter among the different objects can be explained by different star 
formation histories.
On the contrary, the \hDLA\ trend does not seem to agree with the displayed
dust evolution models.
This disagreement is currently debated.
It might be due to the particular history of these systems.
For example, it is possible to have a quasi-linear trend, down 
to low-\hmet, with an episodic star formation history \citep{zhukovska14}.

    \subsubsection{The Aromatic Feature Strength Evolution with Metallicity}
    \label{sec:formPAH}

The \hUIB\ equivalent width clearly rises with metallicity.
It was first demonstrated by \citet{engelbracht05}, using broadband photometry,
and by \citet{madden06}, with spectroscopic observations.
\citet{engelbracht05} argued that there were two regimes: a high \hUIB\
fraction above $\simeq1/5-1/3\;Z_\odot$; and a low value, below.
However, \citet[][hereafter \citetalias{galliano08a}]{galliano08a} demonstrated 
that this was a bias, due to the fact that the continuum dominates the \hMIR\
broadbands when the \hUIB\ strength becomes weak.
Trends constrained with spectroscopy appear rather smooth, although there is
significant scatter (\refpanel{fig:dustevol}{b}).
It was rapidly proposed that the origin of this phenomenon was the enhanced 
destruction of the \hUIB\ carriers in the \hUV\ permeated \hISM\ of 
dwarf galaxies (\cf~\refsec{sec:destPAH}), illustrating again the degeneracy 
between star formation activity and metallicity (\cf~\refsec{sec:SFvsZ}).
Indeed, the \hUIB\ strength appears slightly better correlated with tracers of 
the \hISRF\ hardness than with metallicity \citep[\eg][]{gordon08,wu11}.

\hUIB\ carrier destruction is indubitably an important process in low-\hmet\
systems, but it does not exclude the possible deficiency of their formation.
\citetalias{galliano08a} hypothesized that the \hUIB\ carriers could be 
mostly produced by the long-lived \hAGB\ stars.
This delayed injection mechanism could explain the trend (tracks on 
\refpanel{fig:dustevol}{b}).
However, as we have seen in \refsec{sec:dest}, the \hUIB\ carriers are very 
volatile and need to reform in the \hISM.
Alternatively, \citet{seok14} showed that a dust evolution model in which 
\hPAH s are formed 
by fragmentation of large carbonaceous grains can explain the observed PAH 
trend.
Observationally,
\citet{sandstrom10}, modelling the \hMIR\ spectra of several regions in the
\hSMC, found that the \hPAH\ mass fraction correlates better with the molecular 
gas.
They proposed that \hPAH s could form in molecular clouds.
The trend with metallicity would then result from the lower filling factor of 
the molecular gas at low-\hmet.
This interesting scenario is probably not complete, as the \hPAH\ fractions they 
find in shielded regions are still a factor of $\simeq2-4$ lower than in the \hMW.
This is not surprising as the C/O ratio is about twice lower in the \hSMC\
(\reftab{tab:depletions}).
There is no simple answer to this open question, but it appears that to explain
the paucity of \hUIB s in low-\hmet\ environments, one needs to articulate 
their photodestruction with the deficiency of their formation process.

\begin{issues}[FUTURE CHALLENGES]
  \addcontentsline{toc}{section}{FUTURE CHALLENGES}

\begin{enumerate}
  \item
    Current dust models do not provide a parameterization of the grain mixture 
    constitution as a function of the physical conditions.
    As we have discussed, local evolution processes, like the photodestruction
    of small grains or the mantle growth and evaporation, bias our 
    interpretations, when using models with a fixed constitution.
    Working towards being able to predict, even in a simplified way, the grain 
    properties for an arbitray gas density and \hISRF\ intensity, is 
    necessary to interpret the already existing observations.
  \item
    We are now at a time where nearby galaxies have been observed in detail, 
    over the whole electromagnetic spectrum.
    We are thus compelled to think beyond integrated broadband fluxes to go on 
    progressing.
    \begin{description}
      \item[Spatially resolved studies]
        ($\lesssim1^{\prime\prime}$) are possible in the \hMIR, with \hjwst, 
        and in the \hsubmm, with \hALMA.
        These could help resolve the dust heating in dense extragalactic 
        \hPDR s \refeqp{eq:meanfreepath} and thus provide constraints on the 
        dust properties in high \hISRF\ conditions.
      \item[FIR spectroscopy] (\hspica)
        would be valuable for: 
        \textitem{1}~better constraining the shape of the \hSED; and
        \textitem{2}~identifying new solid state features.
      \item[Multi-process studies]
        are the key to solving the degeneracies inherent to dust models.
        \hluvoir\ could provide valuable constraints on the extinction, 
        depletion,
        \hDIB s and \hERE\ of regions or galaxies which have been observed with
        the previous \hIR--\hsubmm\ telescopes.
    \end{description} 
  \item 
    The increase in sensitivity of the next generation of instruments will 
    be a challenge for data analysis methods.
    More precise fluxes will require proper accounts of foreground and 
    background emissions, especially if we are interested
    in the diffuse \hISM\ of galaxies.
    More complex spatial and spectral decomposition methods, with rigorous 
    treatment of the uncertainties, will be necessary.
\end{enumerate}              
\end{issues}       
\begin{marginnote}
  \entry{JWST}{\hNIR--\hMIR\ space telescope ($\lambda\simeq0.6-27\mmic$;
               launch in 2019),
               with sub-arcsec resolution.}
  \entry{SPICA}{\hMIR--\hFIR\ space telescope ($\lambda\simeq12-210\mmic$; 
                launch in $\simeq2025$), with unprecedented sensitivity.}
  \entry{LUVOIR}{\hUV--\hNIR\ space telescope (in preparation).}
\end{marginnote}

  \section*{DISCLOSURE STATEMENT}

The authors are not aware of any affiliations, memberships, funding, or 
financial holdings that might be perceived as affecting the objectivity of this 
review.

  \section*{ACKNOWLEDGMENTS}

We thank Vincent Guillet for providing us with the polarization model of 
\reffig{fig:dustobs}, Ilse De Looze for \reffig{fig:RT},
Timoth\'e Roland and Ronin Wu for the data of 
\refpanel{fig:MIR}{a}, and Pieter De~Vis for the data in \reffig{fig:scalrel}.
We thank Maarten Baes, Diane Cormier, Pieter De~Vis, Vincent Guillet, Sacha 
Hony, Vianney Lebouteiller, Suzanne Madden, Takashi Onaka and S\'ebastien 
Viaene, for useful discussions and comments, as well as the scientific editor, 
Bruce Draine.
We acknowledge support from the EU FP7 project DustPedia (Grant No.\ 606847).
F.G.\ acknowledges support by the Agence Nationale pour la Recherche 
through the program SYMPATICO (Projet ANR-11-BS56-0023) and
the PRC 1311 between CNRS and JSPS.
M.G.\ acknowledges funding from the European Research Council (ERC) under the 
European Union Horizon 2020 programme (MagneticYSOs project, grant No 679937). 

  \bibliographystyle{ar-style2}
  \bibliography{$HOME/Astro/TeXstyle/references}

\end{document}